\documentclass[paper]{aa} 
\usepackage{graphicx}
\usepackage{natbib}
\usepackage{txfonts}
\pdfoutput=1
\def\Ha{\ifmmode^{\mathrm{H}\alpha }\else$\mathrm{H}\alpha$\fi}
\def\Hb{\ifmmode^{\mathrm{H}\beta }\else$\mathrm{H}\beta$\fi}
\def\LyA{\ifmmode^{\mathrm{H}\alpha }\else$\mathrm{Ly}\alpha$\fi}
\def\BrA{\ifmmode^{\mathrm{Br}\alpha }\else$\mathrm{Br}\alpha$\fi}
\def\BrG{\ifmmode^{\mathrm{Br}\gamma }\else$\mathrm{Br}\gamma$\fi}
\def\PaB{\ifmmode^{\mathrm{Pa}\beta }\else$\mathrm{Pa}\beta$\fi}
\def\mag{\ifmmode^{\rm m }\else$^{\rm m}$\fi}

\def\as{$\,^{\prime\prime}\,$}

\def\hh{\ifmmode^{\rm h}\else$^{\rm h}$\fi}
\def\mm{\ifmmode^{\rm m}\else$^{\rm m}$\fi}
\def\ss{\ifmmode^{\rm s}\else$^{\rm s}$\fi}
\def\deg{\ifmmode^\circ\else$^\circ $\fi}
\def\amin{\ifmmode^\prime\else$^\prime $\fi}

\def\decdm#1#2{\ifmmode{#1}\else{$#1$}\fi\deg\ #2\amin\ }

\def\dec#1#2#3{\ifmmode{#1}\else{$#1$}\fi\deg\ #2\amin\ #3\as\ }

\def\decb#1#2#3#4{\ifmmode{#1}\else{$#1$}\fi\deg\ #2\amin\ #3\farcs#4 }

\begin{document}

\title{
  Revealing the sub-AU asymmetries of the inner dust rim\\
  in the disk around the Herbig~Ae star R\,CrA
  \thanks{Based on observations
    made with ESO telescopes at the La Silla Paranal Observatory under 
    programme IDs 
    079.D-0370(A),
    081.C-0272(A,B,C), and
    081.C-0321(A).
  }
}

\titlerunning{Revealing the sub-AU asymmetries of the inner dust rim in the disk around the Herbig~Ae star R\,CrA}

\author{
  S.~Kraus\inst{1} \and        
  K.-H.~Hofmann\inst{1} \and   
  F.~Malbet\inst{2} \and
  A.~Meilland\inst{3,1} \and
  A.~Natta\inst{4} \and
  D.~Schertl\inst{1} \and      
  P.~Stee\inst{3} \and
  G.~Weigelt\inst{1}
}

\authorrunning{Kraus~et~al.}

\offprints{S.~Kraus}

\institute{
  Max Planck Institut f\"ur Radioastronomie, Auf dem H\"ugel 69, 53121 Bonn, Germany\\
  \email{skraus@mpifr-bonn.mpg.de} \and
  Laboratoire d'Astrophysique de Grenoble, UMR 5571 Universit\'{e} Joseph Fourier/CNRS, BP 53, 38041 Grenoble Cedex 9, France \and
  UMR 6525 H. Fizeau, Univ. Nice Sophia Antipolis, CNRS, Observatoire de la C\^{o}te d'Azur, Av. Copernic, F-06130 Grasse, France \and
  INAF-Osservatorio Astrofisico di Arcetri, Largo Fermi 5, 50125 Firenze, Italy
}

  \date{Received 2009-07-26; accepted 2009-10-28}

 
  \abstract
   {
     Unveiling the structure of the disks around intermediate-mass pre-main-sequence stars (Herbig~Ae/Be stars) 
     is essential for our understanding of the star and planet formation process. 
     In particular, models predict that in the innermost AU around the star, the dust disk forms a ``puffed-up'' inner rim, 
     which should result in a strongly asymmetric brightness distribution for disks seen under intermediate inclination.
   }
   {
     Our aim is to constrain the sub-AU geometry of the inner disk around the Herbig~Ae star R\,CrA
     and search for the predicted asymmetries.
   }
   {
     Using the VLTI/AMBER long-baseline interferometer, we obtained 24 near-infrared ($H$- and $K$-band)
     spectro-interferometric observations on R\,CrA.  
     Observing with three telescopes in a linear array configuration, each data set samples
     three equally spaced points in the visibility function, 
       providing direct information about the radial intensity profile.
     In addition, the observations cover a wide position angle range ($\sim 97${\deg}), also probing
     the position angle dependence of the source brightness distribution.
   }
   {
     In the derived visibility function, we detect the signatures of an 
     extended (Gaussian FWHM $\sim 25$~mas) and a compact component (Gaussian FWHM $\sim 5.8$~mas), 
     with the compact component contributing about two-thirds of the total flux (both in $H$- and $K$-band).
     The brightness distribution is highly asymmetric, as indicated by the 
     strong closure phases (up to $\sim 40${\deg}) and the detected position angle dependence 
     of the visibilities and closure phases.
     To interpret these asymmetries, we employ various geometric as well as physical models, 
     including a binary model, a skewed ring model, and a puffed-up inner rim model 
     with a vertical or curved rim shape.
     For the binary and vertical rim model, no acceptable fits could be obtained.
     On the other hand, the skewed ring model and the curved puffed-up inner rim model 
     allow us to simultaneously reproduce the measured visibilities and closure phases.
     From these models we derive the location of the dust sublimation radius ($\sim 0.4$~AU),
     the disk inclination angle ($\sim 35${\deg}), and a north-southern disk orientation
     (PA$\sim$180-190{\deg}).
     Our curved puffed-up rim model can reasonably well reproduce the 
     interferometric observables and the SED simultaneously and suggests a
     luminosity of $\sim 29~$L$_{\sun}$ and the presence of relatively
     large ($\gtrsim 1.2~\mu$m) Silicate dust grains.
     Our study also reveals significant deviations between the
       measured interferometric observables and the employed puffed-up inner rim models,
       providing important constraints for future refinements of these theoretical models.
       Perpendicular to the disk, two bow shock-like structures appear in the 
       associated reflection nebula NGC\,6729, suggesting that the detected sub-AU 
       size disk is the driving engine of a large-scale outflow.
   }
   {
     Detecting, for the first time, strong non-localized asymmetries in the inner regions of a Herbig~Ae disk, 
     our study supports the existence of a puffed-up inner rim in YSO disks.
   }


\maketitle

%

\section{Introduction}
\label{sec:intro}

For our understanding of the structure and physical processes in the disks around 
young stellar objects (YSOs), the inner-most disk regions
are of special importance.
Furthermore, it is believed that planet formation is a direct result of the grain
aggregation and growth which should take place in the inner few AUs of the 
dusty disks around these stars.
As the spatial scales of the inner circumstellar environment were not 
accessible to imaging observations until recently, most conclusions drawn on the 
geometry of the inner disk were based on the modeling of the spectral energy 
distribution (SED).
Typically, the SED of intermediate-mass YSOs (Herbig~Ae/Be stars) 
shows a characteristic infrared excess emission, which is
often interpreted as the presence of a circumstellar accretion disk, 
although the 3-D geometry of the innermost (AU-scale) region of these disks 
is still poorly known.

According to the current paradigm, most of the infrared excess emission in
Herbig~Ae/Be stars originates from a passive dust disk \citep{ada87}, 
whose thermal structure can be 
approximated with a cold disk interior and a hot surface layer \citep{chi97}.
To fulfill vertical hydrostatic equilibrium, the disk is expected to
flare towards larger radii, allowing the outer disk regions to intercept
more stellar light than expected for a geometrically flat disk \citep{ken87}.
At a certain distance from the star, most often referred to as the dust sublimation
radius $R_{\rm subl}$, the dust temperature will exceed the evaporation temperature 
of dust ($T_{\rm subl}$), causing the truncation of the dust disk and
the formation of a dust-free inner hole.
This scenario also gained support from the first survey-type near-infrared interferometric 
observations of Herbig~Ae/Be stars.
\citet{mil01} could measure the characteristic size
of many YSO disks, finding that the measured sizes scale 
roughly with the square-root of the luminosity $L_{\star}$ of 
the stellar source \citep{mon02}.
Since this is the expected scaling-relation for the location of
the dust-sublimation radius ($R_{\rm subl} \propto L^{1/2}_{\star}$), this finding
supports the idea that the near-infrared emission is tracing mainly
hot material located in a structure at the location of the dust 
sublimation radius.
More recently, significant deviations from the size-luminosity were detected
concerning, in particular, the T~Tauri and Herbig~Be star regime \citep{mon05}.
These deviations might be explained either with contributions from
scattered light \citep[most important for T~Tauri stars; e.g.][]{pin08} or from a gaseous disk 
located inside of the dust sublimation radius
\citep[more important for Herbig~Be stars; see e.g.][]{ake05b,eis07a,kra08a,ise08,tan08}.

In 2001, \citeauthor{nat01} and \citeauthor{dul01} pointed out that 
the heating provided by the stellar radiation should considerably 
increase the disk scale height close to the dust sublimation radius, 
resulting in the formation of a ``puffed-up'' inner rim.  
This rim might cast a shadow on the more extended disk regions, 
possibly affecting the thermal disk structure out to hundreds of AUs.
The shadowing effects of the puffed-up rim might already have been 
observed indirectly in the SED. 
\citet{mee01} argued that the general shape of the SED of 
many Herbig~Ae/Be stars can be divided into two groups, where ``group~I'' sources show a pronounced 60~$\mu$m 
bump, while ``group~II'' sources have a flatter SED, lacking the 60~$\mu$m excess emission.
Later, \citet{dul04a} pointed out that this empirical classification could be explained 
by considering both the flaring properties of the outer disk and the shadowing 
effects of a puffed-up inner rim.
Accordingly, group~I disks might exhibit a strongly flared shape, 
allowing the outer disk to step out of the shadow region, 
while group~II disks are fully self-shadowed.

Although these studies provided some important first insight, 
there are major uncertainties concerning both the magnitude of the puffing-up effect 
as well as the detailed rim shape.
For instance, it was proposed that density-dependent
dust sublimation effects \citep[][referred to as IN05 in the following]{ise05} as well as grain growth and dust
sedimentation \citep{tan07} could affect the shape of the inner rim,
resulting in a curved vertical rim geometry.
In a recent study, \citet{kam09} found that backwarming effects and 
the presence of highly refractory grain species might result in a dust rim 
which is located significantly closer to the star than anticipated in earlier studies.
Using radiative transfer, they show that the rim is not an infinitely
sharp wall, but has an optically thin region which might extend a significant
fraction of the rim radius, resulting in a more diffuse rim morphology.
Spatially resolved observations are required to directly measure the shape 
of the inner rim.
Furthermore, various authors \citep[e.g.][]{mir99} have pointed out that the
infrared emission of many Herbig~Ae/Be stars might contain significant
scattered-light contributions from optically thin circumstellar envelopes or
halos.
As investigated by \citet{vin03} and others, high-angular resolution imaging
observations, such as presented in this study, provide the only method
to separate the disk and envelope contributions and to solve the
ambiguities inherent to pure SED model fits.

One of the strongest predictions which is common to all puffed-up
rim scenarios is the appearance of asymmetries in the source brightness
distribution for disks seen under an intermediate inclination angle.
This ``skew'' in the brightness distribution is a 
direct consequence of the vertical extension of the rim above 
the disk midplane, providing perhaps the most promising way to observationally
distinguish between scenarios with and without a puffed-up rim.
In order to detect these signatures, infrared interferometry 
provides not only the required milli-arcsecond (mas) angular resolution,
but also offers a very sensitive measure for deviations from point-symmetry,
namely the closure phase (CP) relation \citep{jen58}.
First closure phase measurements on YSOs were presented by \citet{mon06}.
Using the IOTA interferometer and baseline lengths up to 38\,m, 
these authors measured statistically significant non-zero closure phases 
on six out of 14 stars (excluding binary stars).  
For five of these six stars, the detected closure phase signals were 
rather small ($\Phi \lesssim 5${\deg}), while for the 
Be star \object{HD\,45677} phases of up to $\sim 27${\deg} could be detected.

In order to investigate the geometry of the inner circumstellar 
environment around a Herbig~Ae star in great detail and to obtain
further evidence for or against the existence of a puffed-up inner rim, 
we have studied the Herbig~Ae star \object{R\,CrA} using 
ESO's Very Large Telescope Interferometer (VLTI) and the AMBER beam 
combination instrument.
In the following, we will first summarize some earlier studies on R\,CrA (Sect.~\ref{sec:rcra}),
followed by a description of our interferometric observations
(Sect.~\ref{sec:observations}) and the employed models
(Sect.~\ref{sec:modeling}).
Finally, we will discuss our modeling results 
(Sect.~\ref{sec:discussion}) and conclude with a brief summary
(Sect.~\ref{sec:conclusions}).

\section{Earlier studies on R\,CrA}
\label{sec:rcra}

R\,CrA is located in the Corona Australis molecular
cloud and is the most luminous \citep[$L_{\rm bol} \sim 99$~L$_{\sun}$,][]{bib92} 
star of the very young and obscured cluster 
known as the Coronet cluster \citep{tay84}.
The stellar parameters of R\,CrA are still rather uncertain,
ranging from a spectral type of F5 \citep{hil92,nat93,gar06}, 
A5 \citep{her88,che97} to B8 \citep{bib92,ham05}.
For the distance, we adopt the commonly assumed value of $130$~pc \citep{mar81}.

Compared to other stars in the Herbig~Ae/Be class, 
R\,CrA is in a particularly early evolutionary phase \citep{mal98b}
and still embedded in an extended and massive natal envelope, 
whose emission dominates the SED from mid-infrared to 
millimeter-wavelengths (Fig.~\ref{fig:modelIN05}, {\it bottom, left}).
\citet{nat93} estimated the mass of the envelope to be 
$\sim 10$~M$_{\sun}$ with an outer envelope radius of 0.007~pc ($\sim 1450$~AU).
In the classification scheme of \citet{mee01}, R\,CrA was classified
as a group~II object \citep{ack04}, suggesting the presence of a disk with a
pronounced inner rim, which might self-shadow the outer disk.
Optical polarization measurements showed a high degree of linear ($\sim 8$\%) and 
circular ($\sim 5$\%) polarization \citep{war85,cla00}, indicating scattering from 
aligned, non-spherical dust grains.
The polarization mapping also showed an extended ($\sim 10$\arcsec), 
disk-like structure with north-south orientation 
\citep[$\theta=189 \pm 5${\deg},][]{war85}.
The total mass of gas and dust in the disk was estimated from sub-millimeter
observations by \citet{man94}, yielding 0.02~M$_{\sun}$.
More recently, \citet{gro07} used the SMA to derive an upper mass limit of $0.012$~M$_{\sun}$.

At optical wavelengths, the star is known to be highly variable both on long 
and short time scales \citep{bel80}.
\citet{gra87} reported strong variability in the H$\alpha$ line of R\,CrA 
as well as surface brightness variations in the nearby reflection nebula NGC\,6729, 
possibly indicating shadowing effects caused by material in the inner
circumstellar environment of R\,CrA or clumpy accretion \citep{gra92}.
Variability was also observed at radio- and X-ray wavelengths \citep{for06}.
The X-ray spectrum of R\,CrA is very unusual, including a very 
hot X-ray emission component.  Intermediate-mass YSOs are
expected to show no X-ray emission, since they should possess neither
magnetically driven coronae nor the radiation-driven winds which cause X-ray-emitting
shock regions.  Therefore, in order to explain the detected X-ray spectrum,
\citet{for06} suggested that the X-ray emission does not originate from the
optical/infrared source, but from a yet undiscovered Class~I companion.
The presence of a close companion was also proposed by \citet{tak03}
based on spectro-astrometric observations which revealed a
photo-center displacement\footnote{Earlier spectro-astrometric observations were presented by \citet{bai98},
but did not show this spectro-astrometric signature.}
both in the blue- and red-shifted wing of the spectrally 
resolved H$\alpha$-line.
The authors pointed out that the detected signature cannot be explained
with a stellar companion or an outflow component alone, but possibly with
a combination of both scenarios.
However, as pointed out by \citet{cho08}, it is unlikely that the 
companion proposed by \citet{tak03} is identical to the Class~I 
companion proposed by \citet{for06}, since a deeply embedded Class~I source would 
not contribute significantly to the H$\alpha$ line flux 
at visual wavelengths.

The accretion activity of R\,CrA was estimated by
\citet{gar06} from the luminosity of the Br$\gamma$-line, 
yielding a relatively low accretion luminosity of $L_{\rm acc}=3.2~L_{\sun}$,
corresponding to a mass accretion rate of $\dot{M}_{\rm acc}=10^{-7.12}~M_{\sun}/{\rm yr}$
(for a F5 star).
Besides indications of active accretion,
various outflow tracers have been reported for R\,CrA.
For instance, a compact bipolar molecular outflow with an east-west orientation 
\citep{wal84,lev88,gra93} as well as several Herbig-Haro objects (in particular
HH\,104~A/B) have been associated with R\,CrA \citep{har87,gra93}.
However, more recent studies \citep{and97,wan04}, convincingly identified
the source IRS\,7 as driving source of these outflows,
making a physical association with R\,CrA rather unlikely.

\section{Observations}
\label{sec:observations}

\begin{figure}[tbp]
  \centering
  \includegraphics[width=8cm]{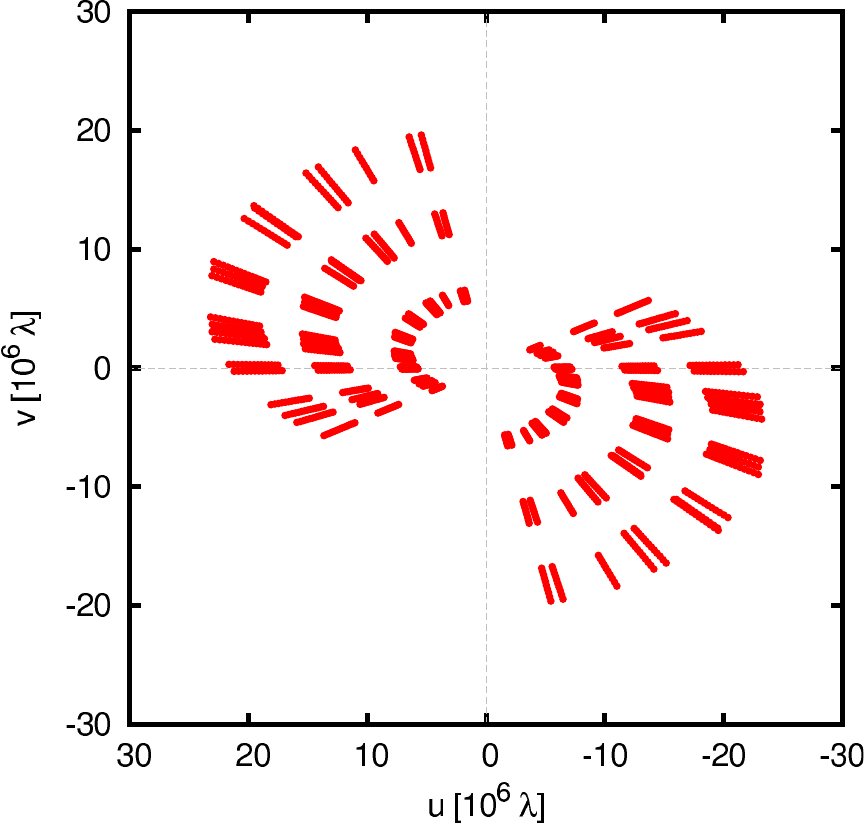}
  \caption{$uv$-plane coverage obtained with our 
    VLTI/AMBER observations on R\,CrA ($K$-band only).
  }
  \label{fig:uvcov}
\end{figure}

\begin{figure*}[tbp]
  \centering
  \includegraphics[width=18cm]{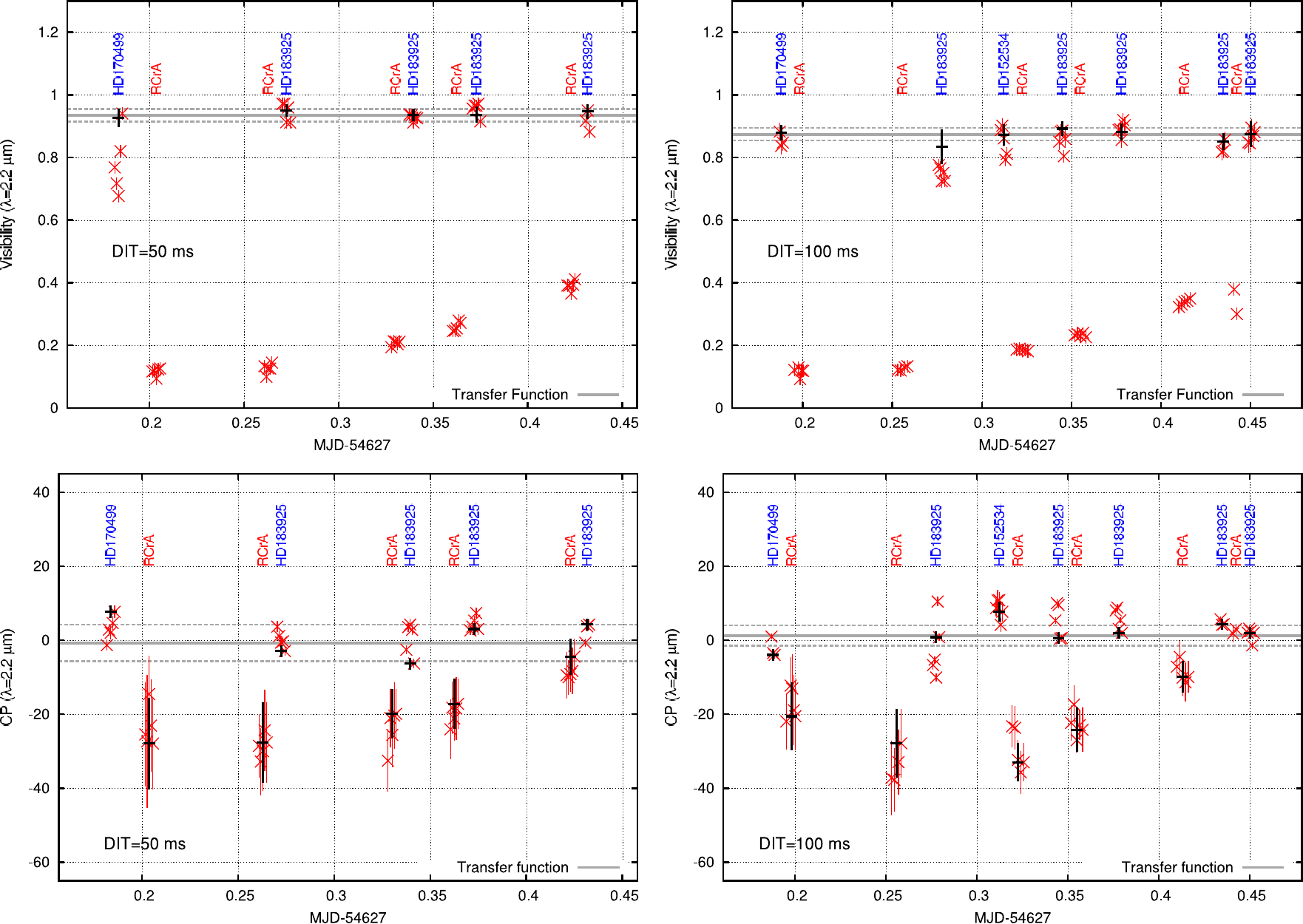}
  \caption{Transfer function of the night 2008-06-09 for one spectral channel around $2.2~\mu$m and
    observations with DITs of 50\,ms {\it (left)} and 100\,ms {\it (right)}.
    The visibilities measured on the longest baseline {\it (top)} and closure phases {\it (bottom)}
    are plotted as a function of time over the night.
    For each block of five exposures (red data points), we compute the average observables
    and correct the calibrator visibilities to correct for their intrinsic UD diameters (black crosses).
    After rejecting data sets with strong intrinsic scatter, 
    we average the calibrator measurements to compute the instrumental transfer function (grey line).
  }
  \label{fig:TFF}
\end{figure*}

\begin{figure}[tbp]
  \centering
  \includegraphics[width=8cm]{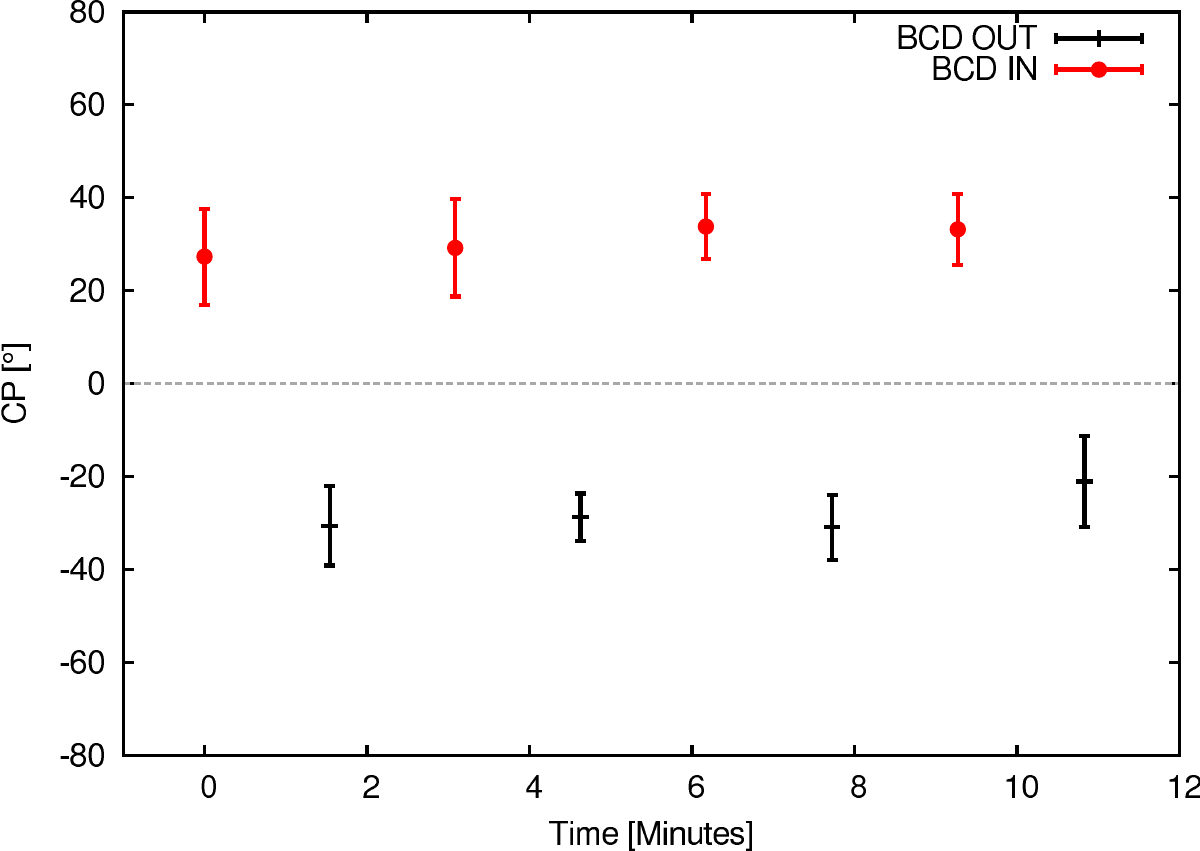}
  \caption{Closure phases recorded on 2008-06-02 on R\,CrA in the $K$-band
    using the AMBER beam commutation device (BCD), showing the
    expected CP sign change between BCD~OUT (black) and BCD~IN (red).
  }
  \label{fig:bcd}
\end{figure}

\begin{figure}[tbp]
  \centering
  \includegraphics[width=8.4cm,angle=0]{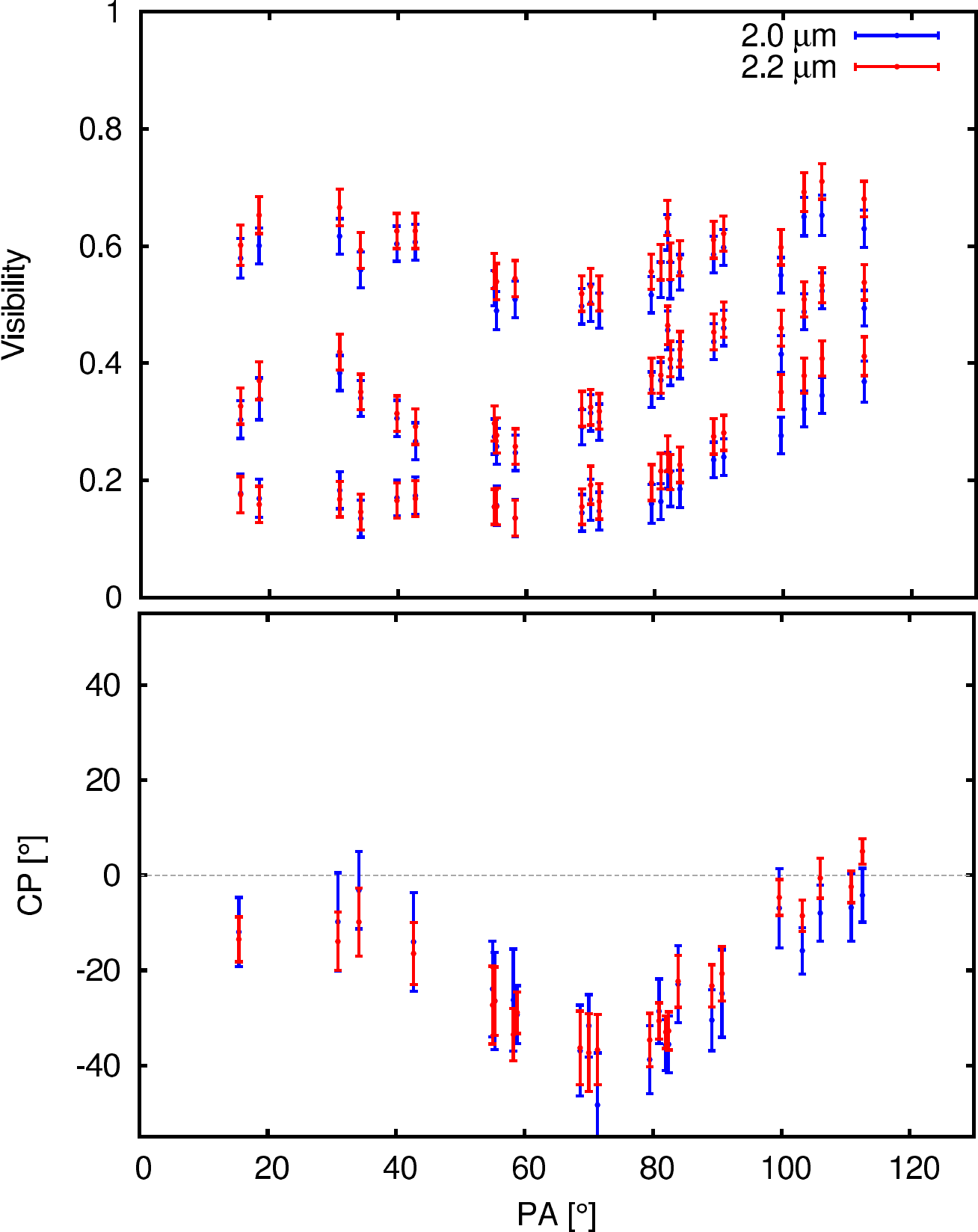} \\
  \caption{Visibilities ({\it top}) and closure phases ({\it bottom}) measured for R\,CrA 
    with the E0-G0-H0 telescope configuration towards different position angles and for 
    two representative spectral bands around $\lambda=2.0~\mu$m ($1.96-2.04~\mu$m) and $2.2~\mu$m ($2.16-2.24~\mu$m).
    In both cases, three spectral channels around the central wavelength were averaged.
  }
  \label{fig:visCP}
\end{figure}

\begin{table}[t]
\caption{Calibrator star information.}
\label{tab:calibrators}
\centering

\begin{tabular}{lccccc}
  \hline\hline
  Star                & $V$  & $K$  & Spectral & $d_{\mathrm{UD}}$\\
                      &      &      & Type     & [mas]           \\
  \noalign{\smallskip}
  \hline
  \noalign{\smallskip}
  \object{HD\,108570} & 6.14 & 4.12 & K1III    & $0.72 \pm 0.05$~$^{(a)}$\\
  \object{HD\,101328} & 7.44 & 3.75 & K4III    & $1.00 \pm 0.01$~$^{(b)}$\\
  \object{HD\,104479} & 4.75 & 3.92 & K0III    & $0.76 \pm 0.01$~$^{(a)}$\\
  \object{HD\,106248} & 6.35 & 3.73 & K2III    & $0.94 \pm 0.01$~$^{(a)}$\\
  \object{HD\,111123} & 1.30 & 1.98 & B0.5IV   & $0.84 \pm 0.06$~$^{(a)}$\\
  \object{HD\,121384} & 6.01 & 4.00 & G8V      & $0.76 \pm 0.05$~$^{(a)}$\\
  \object{HD\,135452} & 6.88 & 3.43 & K3III    & $1.13 \pm 0.02$~$^{(a)}$\\
  \object{HD\,145921} & 6.15 & 3.62 & K2III    & $0.96 \pm 0.01$~$^{(b)}$\\
  \object{HD\,152534} & 6.77 & 4.54 & G8III    & $0.60 \pm 0.04$~$^{(a)}$\\
  \object{HD\,170499} & 7.76 & 3.05 & K4III    & $1.24 \pm 0.02$~$^{(b)}$\\
  \object{HD\,174631} & 6.11 & 2.94 & K1III    & $1.25 \pm 0.02$~$^{(b)}$\\
  \object{HD\,177756} & 3.43 & 3.56 & B9V      & $0.53 \pm 0.04$~$^{(a)}$\\
  \object{HD\,181110} & 7.30 & 3.71 & K3III    & $0.95 \pm 0.01$~$^{(b)}$\\
  \object{HD\,183925} & 6.75 & 2.92 & K5III    & $1.44 \pm 0.02$~$^{(a)}$\\
  \noalign{\smallskip}
  \hline
\end{tabular}

\begin{flushleft}
  \hspace{5mm}{\it Notes}~--~The $V$-band magnitudes were taken from SIMBAD and the $K$-band magnitudes from the 2MASS point source catalog.\\
  \hspace{5mm}{\it References}~--~$(a)$ UD diameter computed with ASPRO.  $(b)$ UD diameter determined with the ESO CalVin tool.\\
\end{flushleft}
\end{table}

We observed R\,CrA during four nights in June 2008 using the VLTI near-infrared
interferometric instrument AMBER \citep{pet07}, which combines the light from three
of the 1.8\,m auxiliary telescopes (ATs).
For all observations, AMBER's low spectral resolution mode (LR-HK) with a
spectral resolution of $\lambda/\Delta\lambda=35$ and a wavelength
coverage from 1.4 to $2.5~\mu$m ($H$- and $K$-band) was used.
The telescopes were placed on stations E0-G0-H0, forming a linear array
configuration with baseline lengths of 16 (E0-G0), 32 (H0-G0), and 48\,m (E0-H0).
Due to this linear telescope arrangement 
and the spectral coverage provided by the AMBER instrument, each of our AMBER 
observations provides a good sampling of the $uv$-plane 
towards a certain position angle (PA), covering 
spatial frequencies between $\sim 4$ and $25 \times 10^{6}\lambda$, 
allowing one to cover a wide range of the visibility function 
($0.8 \lesssim V \lesssim 0.1$).
By obtaining a large number of observations towards different hour angles,
we also achieved a good PA coverage, probing PAs between 16{\deg} and 113{\deg}
(Fig.~\ref{fig:uvcov}).
In order to adjust for changing atmospheric conditions and to test
for possible systematic effects, we recorded data with different detector integration 
times (DITs), namely 50~ms (default value), 100~ms, and 200~ms.

The science observations on R\,CrA were interlayed with observations on
interferometric calibrator stars, allowing us to monitor the instrumental 
transfer function.  The transfer function is used in the course of data 
reduction to correct for wavelength-dependent instrumental or atmospheric effects.
Each observation block (either on R\,CrA or on the calibrator star)
typically consisted of five data sets, each containing   
1000 (for DIT=50~ms) or 500 (for DIT=100~ms or DIT=200~ms) individual
spectrally dispersed interferograms.

For data reduction, we used the {\it amdlib} data reduction software (release 2.2).
This software employs the P2VM algorithm \citep{tat07b} to derive 
wavelength-dependent visibilities and closure phases (CP).
The wavelength calibration was done using the procedure described in
Appendix~A of \citet{kra09}.
Following the standard AMBER data reduction procedure, we select the 10\% of interferograms
with the best signal-to-noise (SNR) ratio.
Furthermore, we reject interferograms which were taken with an optical path delay (OPD)
larger than 4~$\mu$m in order to avoid systematic degenerative effects.
To determine the transfer function, we first plot the observables
for a representative spectral channel as a function of time (Fig.~\ref{fig:TFF}).
Then, we reject observation blocks which show a significant intrinsic scatter 
between the five individual exposures, typically indicating some degraded quality
due to poor atmospheric conditions.
While very few observations have to be rejected for the $K$-band,
only a few $H$-band measurements provide reliable results,
reflecting the lower brightness of the object in this spectral window
and the lower stability of the atmosphere at shorter wavelengths.
Finally, we correct for the intrinsic diameters of the 
calibrator stars used (Tab.~\ref{tab:calibrators}) and average the diameter-corrected calibrator 
observations for each night and for each spectral channel in order to yield the transfer function.
The calibration error is estimated from the scatter of the individual
calibrator observations over the night.
As can be seen in Fig.~\ref{fig:TFF}, the calibration uncertainty is typically about 
$\sim 3$\% for visibility and $\sim 5^{\circ}$ for CPs.
Besides this calibration error, we include the statistical error estimate 
provided by {\it amdlib}.
Since this procedure results in unrealistically small error bars for some individual measurements, 
we add a constant calibration uncertainty of $3$\% for the visibility measurements.
As can be seen in Fig.~\ref{fig:TFF}, the measurements obtained 
with different DIT values agree very well with each other, indicating good data 
quality.

For some of our observations on R\,CrA, we employed the
AMBER beam commutation device \citep[BCD,][]{pet07}.  
This calibration device allows one to exchange the beams of two of the three 
telescopes within a few seconds and to trace potential drifts
in the CP transfer function on much shorter time scales than a typical
VLTI target/object cycle ($\sim 30$~min).
Furthermore, the BCD device is located close to the beginning of the optical path of
the AMBER instrument and inverts the phase sign of the two exchanged
beams, while the instrument-internal phases are not affected \citep{mil08}.
Due to this optically introduced change in the CP sign, it is possible 
to distinguish real astrophysical CP signatures from 
potential systematic instrumental artifacts.
Our BCD observations on R\,CrA were performed on 2008-06-02
and show the expected change of sign (Fig.~\ref{fig:bcd}),
confirming the astrophysical origin of the detected CP signals.

\begin{figure*}[tbp]
  \centering
  \includegraphics[width=18cm]{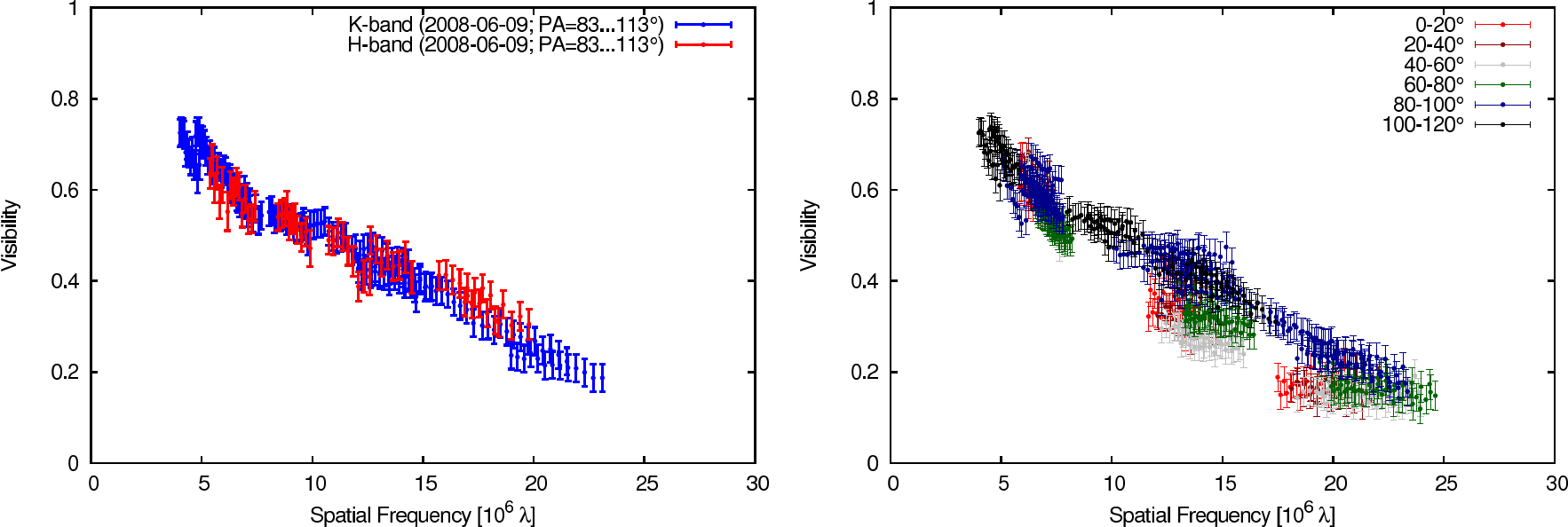}
  \caption{
    {\it Left:} Visibilities derived from four R\,CrA observations taken on 
    2008-06-09 and covering the PA range between 83{\deg} and 113{\deg}. 
    For the $H$- and $K$-band spectral channels covered by our observations, 
    the measured visibilities follow practically an identical visibility profile.
    {\it Right:} Comparing the visibility measured during four observing nights towards very 
    different position angles (coded with different colors) reveals a strong position angle 
    dependence of the visibility function, in particular at spatial frequencies above $8 \times 10^{6} \lambda$.
  }
  \label{fig:visHKvisPA}
\end{figure*}

\begin{figure*}[tbp]
  \centering
  \includegraphics[width=18cm]{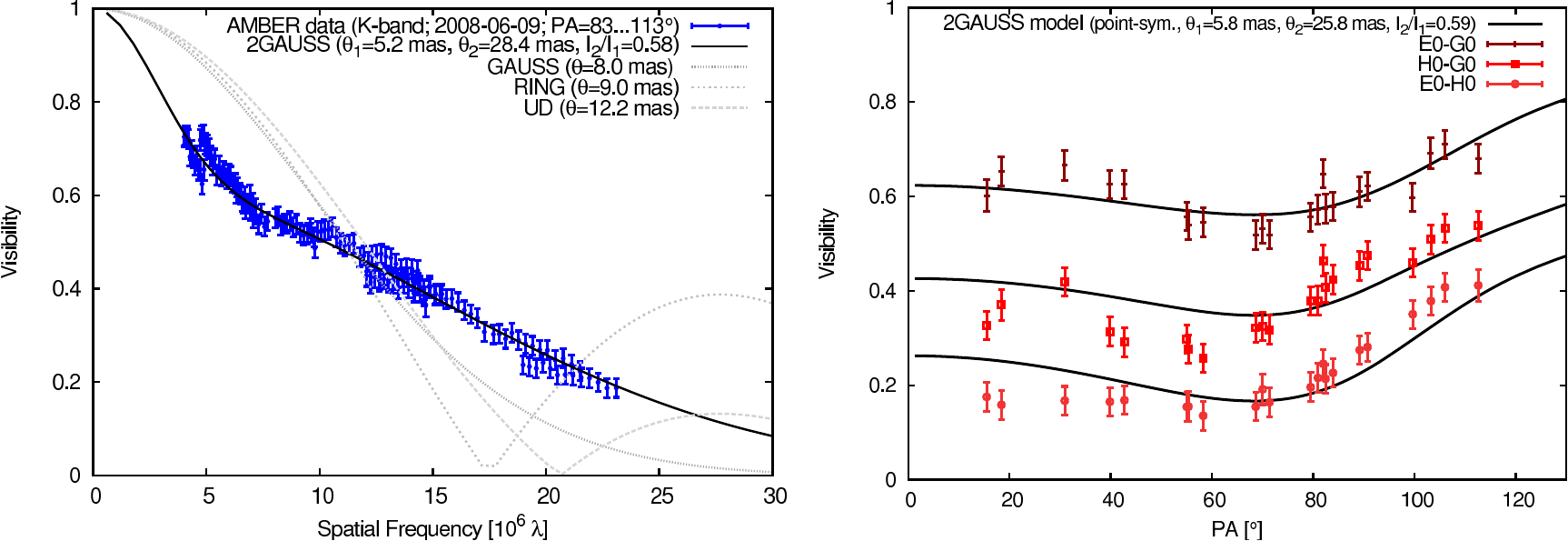}
  \caption{ 
    {\it Left:} Comparison of the $K$-band visibilities measured for the 
    PA range 83-113{\deg} with various point-symmetric model geometries.
    It is evident that one-component models, such as UD, RING, and GAUSS geometries, 
    are not consistent with the visibility profile, while a
    point-symmetric two-component Gaussian model (2-GAUSS) can reproduce the visibilities 
    reasonably well, possibly indicating the 
    presence of two spatial components such as an envelope and a disk (Sect.~\ref{sec:model2GAUSS}).
    {\it Right:} Fitting the 2-GAUSS model to the visibility data measured towards
    very different PAs shows that this model cannot reproduce the measured position angle 
    dependence of the visibilities, in particular on the longer baselines.
   }
  \label{fig:vis2GAUSSSYM}
\end{figure*}

Fig.~\ref{fig:visCP} shows the position angle dependence of the derived 
visibilities and CPs for two spectral windows around 2.0 and 2.2~$\mu$m.
In order to interpret these position angle dependent variations, it is necessary
to model the object morphology, taking also the variations of the projected baseline
length with position angle into account, as will be done in Sect.~\ref{sec:modeling}.
In Fig.~\ref{fig:visHKvisPA} ({\it left}), we plot the visibilities measured for the PA range
83-113{\deg} as a function of spatial frequency.
The spatial frequency $\nu=B/\lambda$ provides a measure for the resolving power achieved with a certain observation
and is therefore proportional to the projected baseline length and inverse proportional to the observing wavelength $\lambda$.
As can be seen in the left panel, the visibilities measured for different spectral channels 
in the $H$- and $K$-band can be very well represented with the same visibility profile.
This indicates that the brightness distribution does not show a strong
wavelength dependence\footnote{This argument is valid only 
for the visibility function $V$, but might not apply to the 
wavelength dependence of the closure phases, which are more sensitive to 
small-order changes in the object morphology
\citep[$V \propto \nu^2$, $\Phi \propto \nu^3$; see][]{lac03}.}, 
which has interesting implications on the temperature-distribution
of the emitting physical structure (see discussion in Sect.~\ref{sec:discussionRimGeometry}).
Given the relatively small number of reliable $H$-band 
measurements, we consider only the $K$-band measurements for the following 
quantitative modeling.
Also, the measured $H$-band closure phases are associated with 
very large error bars and are therefore omitted for the quantitative analysis.

\section{Modeling}
\label{sec:modeling}

\begin{table*}[t]
\caption{Model-fitting results.}
\label{tab:modelfitting}
\centering
\begin{tabular}{l|ccccccccc|cc|ccc}
  \hline\hline
                & \multicolumn{9}{c|}{Compact component}  & \multicolumn{2}{c|}{Extended component} & \multicolumn{3}{c}{Goodness-of-fit} \\
  \hline
  Model         & $\theta_{1}$ & $f$          & $\rho$ & $\phi^{(a)}$ & $i$    & $s$ & $R$     &$H/R$   & $\epsilon$ & $\theta_{\rm 2}$& $I_{\rm 2}/I_{\rm 1}$  &  $\chi^{2}_{\rm r,V}$ & $\chi^{2}_{\rm r,\Phi}$ & $\chi^{2}_{\rm r}$ \\
                & [mas]        &              & [mas]  & [\deg]      & [\deg] &     & [AU]   &        &            & [mas]          &                 &                     &                       & \\
  \hline
  \noalign{\smallskip}
                & \multicolumn{9}{c|}{\bf{Point-symmetric models:}} & & & & & \\
  UD            & 13.86        &              &        &             &       &       &        &        &           &                 &                & 24.82 & $29.32^{(c)}$ & 25.92 \\
  RING          & 8.92         & $0.25^{(b)}$  &        &             &       &       &        &        &           &                 &                & 26.38 & $29.32^{(c)}$ & 25.21 \\
  GAUSS         & 8.55         &              &        &             &       &       &        &        &           &                  &               & 18.16 & $29.32^{(c)}$ & 20.88 \\
  2-GAUSS       & 5.77         &              &        &             &       &       &        &        &           & 25.78            & 0.59          & 2.35 & $29.32^{(c)}$ & 8.92 \\
                & \multicolumn{9}{c|}{\bf{Asymmetric models:}} & & & & & \\
  BINARY        & $4.8$        &              & $6.2$  & $34$        &       &       &        &        &           & $14.8$           & $1.04$        & 3.47 & 5.32        & 3.93 \\  
  SKEWED RING   &              & $0.8$        &        & $190$       & $14$  & $0.64$& $0.44$ &        &           & $27$             & $0.58$        & 1.58 & 3.13        & 1.96 \\  
  VERTICAL RIM  &              &              &        & $132$       & $16$  &       & $0.60$ & $0.35$ &           & $27$             & $0.68$        & 2.30 & 6.16        & 3.25 \\
  CURVED RIM    &              &              &        & $180$       & $35$  &       &        &        & $\epsilon_{\rm cr}$ & $32$             & $0.50$        & 1.77 & 3.28        & 2.14 \\
  \noalign{\smallskip}
  \hline
\end{tabular}
\begin{flushleft}
  \hspace{5mm}{\it Notes}~--~
  $(a)$~This column gives the model orientation, measured East of North.  
  For the BINARY model, $\phi$ gives the PA of the separation vector, while for the SKEWED RING, 
  VERTICAL RIM, and CURVED RIM models, the orientation of the ellipse/rim major axis is given.
  Due to a lack of closure phase calibration observations on the E0-G0-H0 array configuration, 
  we are unfortunately not able to unambiguously define the CP sign, resulting
  in a 180{\deg}-ambiguity in the derived position angles.
  $(b)$~In the fitting process, this parameter was fixed. 
  $(c)$~This model is point-symmetric, resulting in a CP which is identical zero.\\
\end{flushleft}
\end{table*}

In this section we describe the geometric and physical models which we 
fitted to our interferometric data to constrain the spatial 
distribution of the circumstellar material around R\,CrA.
Besides circumstellar dust and gas emission, the $K$-band might 
contain flux contributions from the stellar photosphere.
Therefore we estimate the photospheric flux contributions by comparing 
the de-reddened SED of R\,CrA with atmospheric models (Fig.~\ref{fig:modelIN05}), 
yielding relatively small values between $\sim0.5$\% \citep[assuming the spectral type F5,][]{gar06} 
and $\sim3$\% \citep[spectral type B8,][]{bib92} of the total flux
for the $K$-band.
Given this result, it seems justified to neglect the photospheric 
contributions and to include only the thermal emission of hot circumstellar 
material in our modeling process (with the exception of the
curved rim model, see Sect.~\ref{sec:modelIN05}).

In order to fit the described models to our interferometric data,
we first generate model images for each model and each set of parameters.
By construction, the geometric models (Sects.~\ref{sec:modelspherical} to \ref{sec:modelVERTRIM}) 
are monochromatic, while for our curved rim model (Sect.~\ref{sec:modelIN05}), 
we compute the brightness distribution for each spectral channel separately, 
assuming the computed surface layer temperature for each disk annulus.
From these model images (with a scale of 0.2~mas/pixel), 
we compute the Fourier amplitudes (visibilities) and Fourier phases for the 
$uv$-coordinates covered by the interferometric observations.
Model visibilities $V^{\prime}$ and closure phases $\Phi^{\prime}$ are computed 
for each spectral channel separately and then compared to the 
observables ($V$, $\Phi$) and their uncertainties ($\sigma_{V}$, $\sigma_{\Phi}$)
by adopting $\chi_r^{2} = \chi^{2}_{\rm r,V} + \chi^{2}_{\rm r,\Phi}$ as the likelihood estimator, with
\begin{eqnarray}
  \chi^{2}_{\rm r,V}   & = & \frac{1}{N_{V}} \sum_{i=1 \ldots N} \left[ \left( \frac{V_{i} - V^{\prime}_{i}}{\sigma_{V_{i}}} \right)^{2} \right] \\
  \chi^{2}_{\rm r,\Phi} & = & \frac{1}{N_{\Phi}} \sum_{i=1 \ldots N} \left[ \left( \frac{\Phi_{i} - \Phi^{\prime}_{i}}{\sigma_{\Phi_{i}}} \right)^{2} \right],
\end{eqnarray}
where $N_{V}$ and $N_{\Phi}$ are the degrees of freedom for the
individual visibility and closure phase measurements, respectively.
In order to find the best-fit set of parameters for each model,
we vary all free parameters on a parameter grid and search for 
the global $\chi_r^{2}$ minimum.

In the following, we give a detailed description of the
applied models and show the obtained best-fit results for the
spectral channel around 2.2~$\mu$m (Figs.~\ref{fig:visHKvisPA} to \ref{fig:modelIN05}). 
A comparison between the model and the data obtained in all spectral channels is shown
in Figs.~\ref{fig:overviewVIS} and \ref{fig:overviewCP}.
The derived best-fit parameters are given in Tab.~\ref{tab:modelfitting}.

\subsection{UD, RING, GAUSS: Point-symmetric models}
\label{sec:modelspherical}

For a first quantitative interpretation of our data, we employed the
commonly used uniform disk (UD), ring (RING), and Gaussian (GAUSS) geometries.
By construction, these simple geometries are point-symmetric
and, thus, not able to reproduce any object elongation or object asymmetries.
Therefore, in contrast to the following Sects.~\ref{sec:modelbinary} to
\ref{sec:modelIN05}, here we aim only at reproducing the 
measured visibilities, being aware that the real object geometry is, 
in fact, not point-symmetric. 
Nevertheless, these simple models can be useful to estimate the 
characteristic size $\theta$ of the source brightness distribution and to allow 
a comparison with interferometric studies of other objects.
In order to minimize potential influences from object elongation, 
we applied these model fits to a data subset, covering a smaller PA range.
Due to the large number of available independent measurements, we
have selected the PA interval 83-113{\deg}.
From the obtained best-fits (grey curves in Fig.~\ref{fig:vis2GAUSSSYM}, {\it left})
it is evident that neither of these point-symmetric standard geometries can reproduce 
the measured visibilities.

\subsection{2-GAUSS: Indications for a disk+envelope geometry}
\label{sec:model2GAUSS}

A plausible explanation for the detected strong deviations between 
standard model geometries and the measured visibilities might be the
presence of multiple components, such as a disk and an envelope component.
This scenario also gains support from earlier studies which required 
the presence of a particularly massive circumstellar envelope \citep{nat93}
to reproduce the SED of R\,CrA.

As a simple two-component geometry, we considered a model which consists of 
two Gaussian components.
In a first attempt (Model 2-GAUSS), we fixed the center of the two 
Gaussians and varied only their FWHM diameter ($\theta_{1}$, $\theta_{2}$)
and intensity ratio ($I_{2}/I_{1}$),
which allows us to reproduce the visibility profile 
measured towards a narrow PA-range reasonably well (Fig.~\ref{fig:vis2GAUSSSYM}, {\it left}).
The parameters of this two-Gaussian model (model 2-GAUSS)
are listed in Tab.~\ref{tab:modelfitting}.
Of course, being point-symmetric, the model 
cannot reproduce the measured non-zero closure phases.
However, the model allows one to explain the pronounced change of slope, 
which can be seen in the visibility
function around spatial frequency $\nu \sim 8 \times 10^{6} \lambda$.
At shorter spatial frequencies, the measured visibilities do not show any 
significant position angle dependence, while at higher spatial frequencies, 
the visibilities differ significantly toward different position angles
(see Fig.~\ref{fig:visHKvisPA}, {\it right}).
This strongly suggests the presence of at least two spatial components,
one being rather extended and point-symmetric (dominating at $\nu \lesssim 8 \times 10^{6} \lambda$)
and the other being more compact and strongly asymmetric (and dominating at $\nu \gtrsim 8 \times 10^{6} \lambda$).
Based on this rather general argument, we include two spatial components
in all model geometries, namely
\begin{itemize}
\item[{\it (a)}] an extended, symmetric component, which we represent with a Gaussian 
  (in the following, this component will be referred to as \textit{``Envelope''}) and
\item[{\it (b)}] a compact, elongated, and asymmetric component (in the following referred 
  to as \textit{``Disk''}), which has the purpose of reproducing the measured non-zero 
  closure phases and the detected position angle dependence of the interferometric observables.
\end{itemize}
Of course, for our modeling, the two components only represent geometric 
quantities on different spatial scales, and their denotion as ``Envelope'' and ``Disk'' 
is, at the modeling stage, done only for the sake of clarity.
Therefore, we include in all following models 
(with the exception of the 2-GAUSS and BINARY model, which consist of two Gaussian components itself)
an envelope component, which is represented by two free parameters, namely
the FWHM diameter ($\theta_{2}$) and the envelope/disk flux ratio ($I_{2}/I_{1}$).

\subsection{BINARY: Asymmetric two-component Gaussian model}
\label{sec:modelbinary}

\begin{figure}[tbp]
  \centering
  \includegraphics[width=9cm]{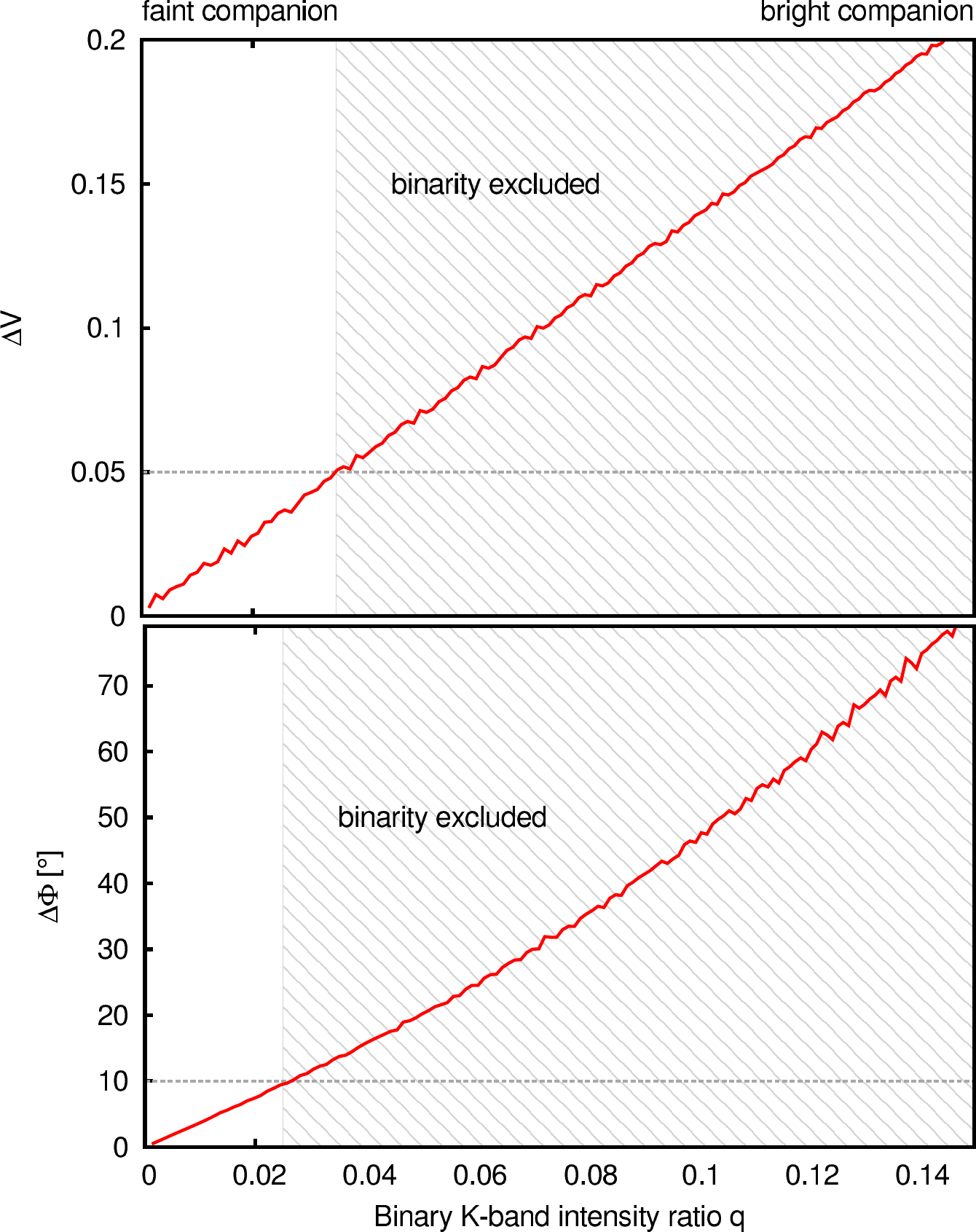}
  \caption{
    Simulation of the expected visibility ({\it top}) and 
    closure phase ({\it bottom}) signatures of a
    hypothetical companion star, plotted as function of the
    binary flux ratio.
    For these simulations, we assume a point-symmetric 
    extended component (for which we use the 2-GAUSS geometry
    discussed in Sect.~\ref{sec:model2GAUSS}) and add  
    a compact companion at a separation of 60~mas,
    corresponding to the minimum separation of the 
    proposed wide-separation companion star \citep{tak03}.
    Then, we compute the residuals between the model 
    with and without companion star 
    and measure the amplitude of the wavelength-differential
    visibility and CP
    modulation for the $K$-band spectral window.
    The dashed horizontal lines give the achieved 
    wavelength-differential visibility and closure phase accuracy, 
    indicating that our observations should be sensitive to
    any companion star contributing more than $q=1:40$ of the 
    total flux (shaded area).
  }
  \label{fig:simbinary}
\end{figure}

\begin{figure}[tbp]
  \centering
  \includegraphics[width=9cm]{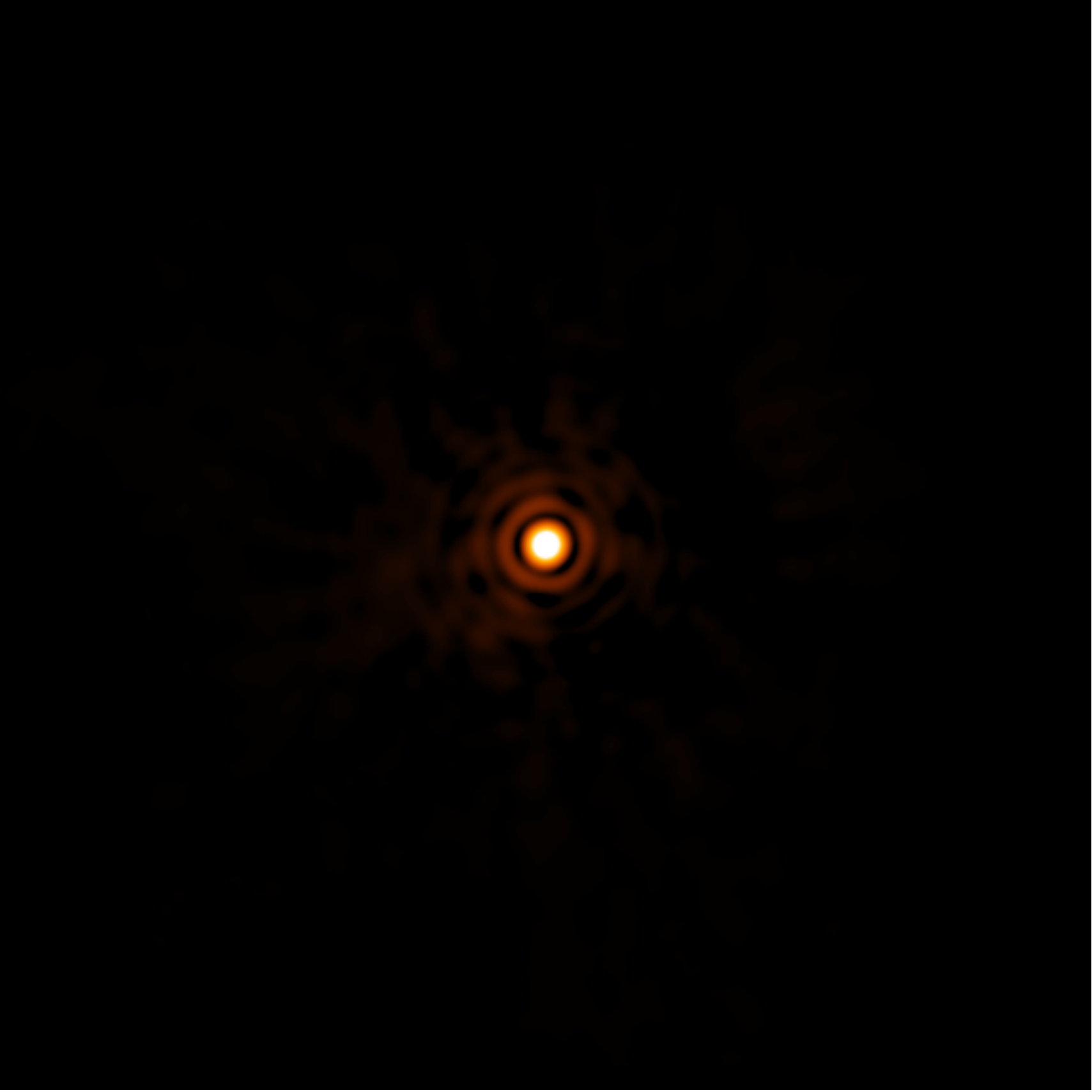}
  \caption{
    Bispectrum speckle image reconstructed from $H$-band speckle interferograms recorded with the 
    ESO\,3.6m telescope on 2007-04-24 (field-of-view 6\arcsec). 
    At the achieved diffraction-limited angular resolution of 95\,mas, 
    R\,CrA appears as a point-source.  
    The first and second diffraction ring are clearly visible with 
    noise-features on the 1\%-level, allowing us to rule out 
    the existence of a companion star in this separation/brightness contrast range.
  }
  \label{fig:speckle}
\end{figure}

\begin{figure*}[t]
  \centering
  \includegraphics[width=18cm]{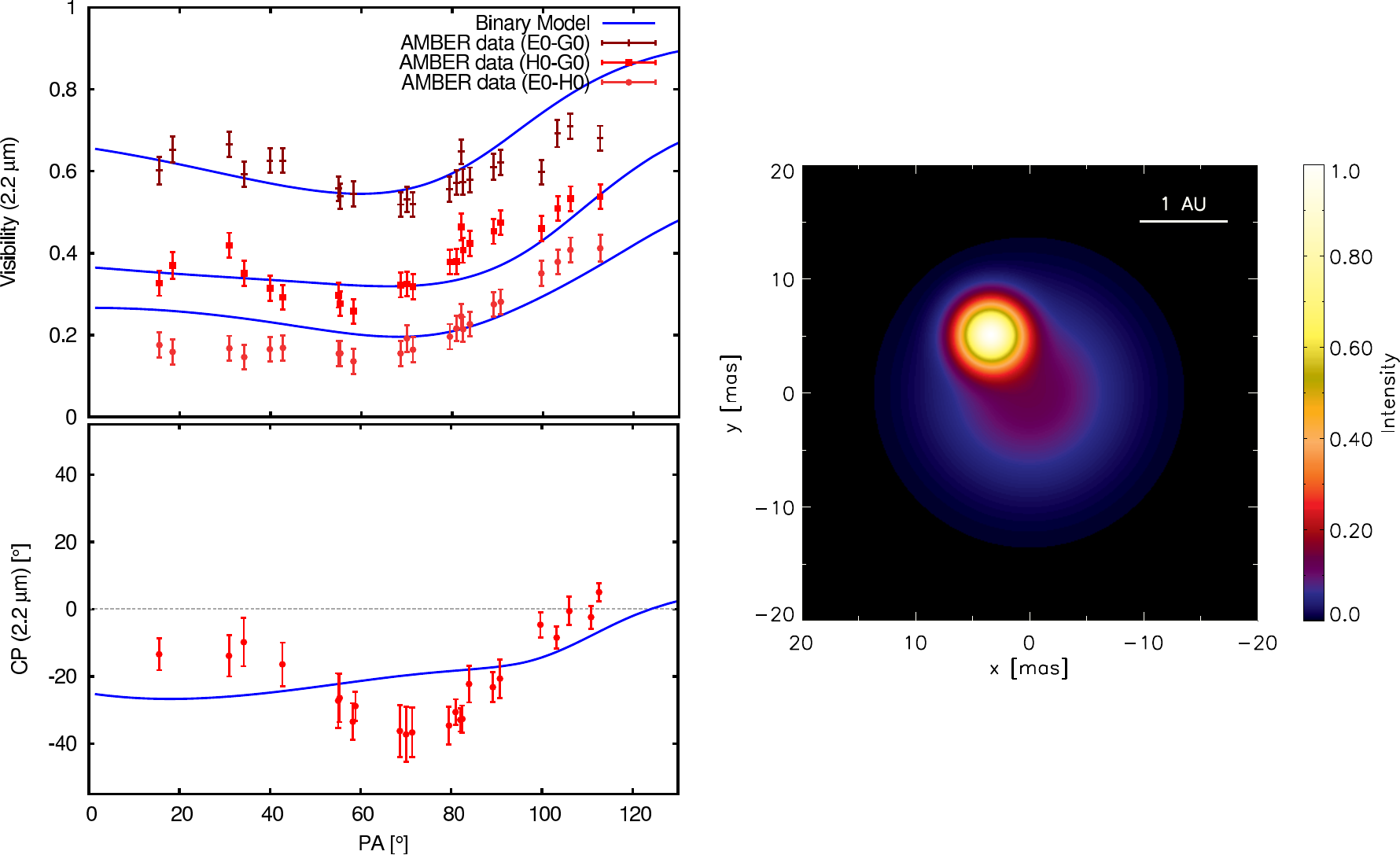}
  \caption{
    {\it Left:} Comparison between the AMBER observables and the model 
    visibilities ({\it top}) and closure phases ({\it bottom}) from
    our BINARY model.
    {\it Right:} Image corresponding to our best-fit model.
  }
  \label{fig:modelBINARY}
\end{figure*}

The presence of close companions can introduce strong asymmetries
in the source brightness distribution and might cause strong
non-zero closure phases such as those detected in our VLTI/AMBER measurements.
For R\,CrA, this interpretation seems particularly appealing, since 
two earlier studies proposed the presence of multiple stellar sources
to explain the measured spectro-polarimetric signatures \citep{tak03} 
and the unexpected hard X-ray emission \citep{for06}.
From their spectro-astrometric signal, \citet{tak03} derived 
a lower limit of 8~AU ($ \gtrsim 60$~mas) for the apparent separation of 
this hypothetical companion.
Given this wide separation, the binary should be clearly detectable
in our AMBER data.
As illustrated in Fig.~3 of \citet{kra09}, binaries with such a wide separation
cause a high-frequency cosine modulation in the wavelength-differential 
visibilities and closure phases 
(this argument holds as long as the separation is large
compared to the diameter of the individual components).
Neither the measured visibilities nor closure phases show such a 
systematic, wavelength-differential modulation (Fig.~\ref{fig:overviewVIS} 
and \ref{fig:overviewCP}), which clearly indicates that no companion
with a separation $\rho \gtrsim 20$~mas 
is significantly contributing to the near-infrared emission.
In order to investigate up to which flux ratio $q=I_{\rm comp}/I_{\rm total}$ 
our observations would be sensitive to the presence of such a hypothetical 
companion star, we simulated the $K$-band wavelength-differential 
signatures of a compact companion star around a point-symmetric, extended 
component such as derived in Sect.~\ref{sec:model2GAUSS} and compare the 
amplitude of the predicted wavelength-differential modulations with the achieved 
differential visibility and CP accuracy.
Based on these simulations (Fig.~\ref{fig:simbinary}), 
we estimate that our observations rule out 
the existence of a companion star for the 
separation range 20~mas~$\lesssim \rho \lesssim$~200~mas 
and for $K$-band flux ratios brighter than $\sim$1:40.
The upper limit in this separation range is given by the field-of-view
of the used VLTI/AT telescopes. 
Besides these constraints from VLTI/AMBER interferometry,
we obtained bispectrum-speckle interferometric 
observations with the ESO\,3.6m telescope ($H$-band; data from 2007-04-24 and 2007-04-27)
using the Rockwell HAWAII detector of our visitor speckle camera.
For image reconstruction we used the bispectrum speckle interferometry method
(\citealt{wei77}, \citealt{wei83}, \citealt{loh83}).
Both the derived power-spectrum and the reconstructed
diffraction-limited bispectrum speckle image (Fig.~\ref{fig:speckle}),
shows R\,CrA as a point-source, ruling out the existence of a companion star 
at separations $\rho \gtrsim 60$~mas, down to a flux ratio of 1:40.

In order to test for the presence of a companion at shorter
separation, we also fitted a binary star model to our AMBER data.
Since it is likely that the near-infrared emission of the two components 
is not dominated by photospheric emission, but instead by the thermal emission of 
hot circumstellar material, we represent the two components not with point-sources, 
but with extended geometries, namely Gaussians.
The choice of this geometry is motivated by our results from Sect.~\ref{sec:model2GAUSS}, 
where we showed that two Gaussians are well suited to represent the
radial intensity profile towards some PA ranges.

In this model, the free parameters are the FWHM sizes of the two Gaussians 
($\theta_{1}$, $\theta_{2}$), their flux ratio ($I_{2}/I_{1}$), 
their angular separation ($\rho$), and the position angle ($\phi$).
As shown in Fig.~\ref{fig:modelBINARY}, the best-fit binary model can provide
only a moderate fit to the measured visibilities and CPs ($\chi_r^{2}=3.93$).
Furthermore, as will be discussed in Sect.~\ref{sec:discussionNoBinary}, the 
found best-fit solution is not physically meaningful in the context of a companion star
scenario.

\subsection{SKEWED RING model}
\label{sec:modelskewedring}

\begin{figure*}[t]
  \centering
  \includegraphics[width=18cm]{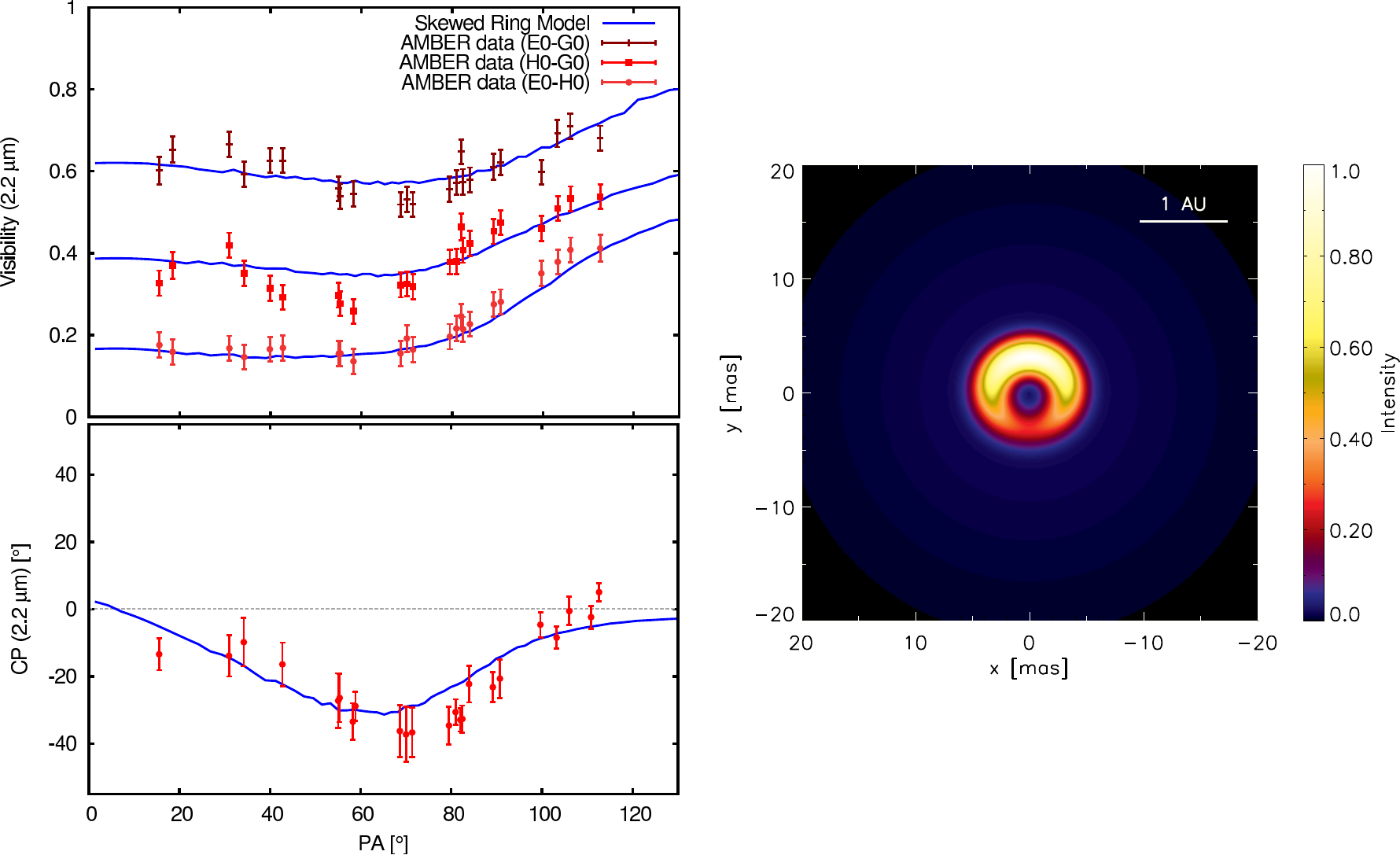}
  \caption{
    {\it Left:} Comparison between the AMBER observables and the model 
    visibilities ({\it top}) and closure phases ({\it bottom}) from
    our SKEWED RING model.
    {\it Right:} Image corresponding to our best-fit model.
  }
  \label{fig:modelSKEWEDRING}
\end{figure*}

For passive irradiated disks, most of the near-infrared emission 
should emerge from hot dust located in a narrow annulus around the dust 
sublimation radius.
Therefore, \citet{mon02} and others have argued that ring geometries
might provide a good approximation for the appearance of YSO disks on AU-scales.
While for face-on disks, point-symmetric rings seem appropriate, ring models are 
likely to be an over-simplification for disks seen under significant inclination.
Since one side of the rim should appear brighter than the other, 
\cite{mon06} proposed a simple mathematical modification of the standard ring model using a
sinusoidal modulation of the ring brightness as a function of the azimuthal angle.
We apply a similar modeling approach using the intensity distribution
\begin{equation}
  I(r, \alpha) = \left(1 - s \cdot \sin(\alpha - \phi) \right) \cdot \exp \left( - \frac{4 \ln 2 \cdot (r-p R)^2}{(f p R)^2} \right), 
\end{equation}
where the first term modulates the brightness distribution as a Gaussian centered at radius 
$R$ and of fraction width $f$ (FWHM).
The brightness distribution $I(r, \alpha)$ is defined in a polar coordinate
system, where $r$ denotes the radius from the center and $\alpha$ the azimuth angle.
The projection factor $p$ can be used to take inclination effects into account and
is given by
\begin{equation}
p = \cos i \left( \cos^2 i \cdot \cos^2(\alpha-\phi) + \sin^2(\alpha-\phi) \right)^{-1/2}.
\end{equation}
In both equations, $\phi$ denotes the PA along which the major axis 
of the apparent disk ellipse is aligned.
The fitted free parameters are the ring radius ($R$), the fractional ring width
$f$, the position angle $\phi$, the inclination $i$, the skew parameter $s$, as well
as the two envelope parameters ($\theta_{2}$, $I_{2}/I_{1}$) described above.
For this skewed ring model, our fitting algorithm can reach good agreement
with the data ($\chi_r^2=1.96$; Fig.~\ref{fig:modelSKEWEDRING}).

\subsection{VERTICAL RIM model}
\label{sec:modelVERTRIM}

\begin{figure*}[t]
  \centering
  \includegraphics[width=18cm]{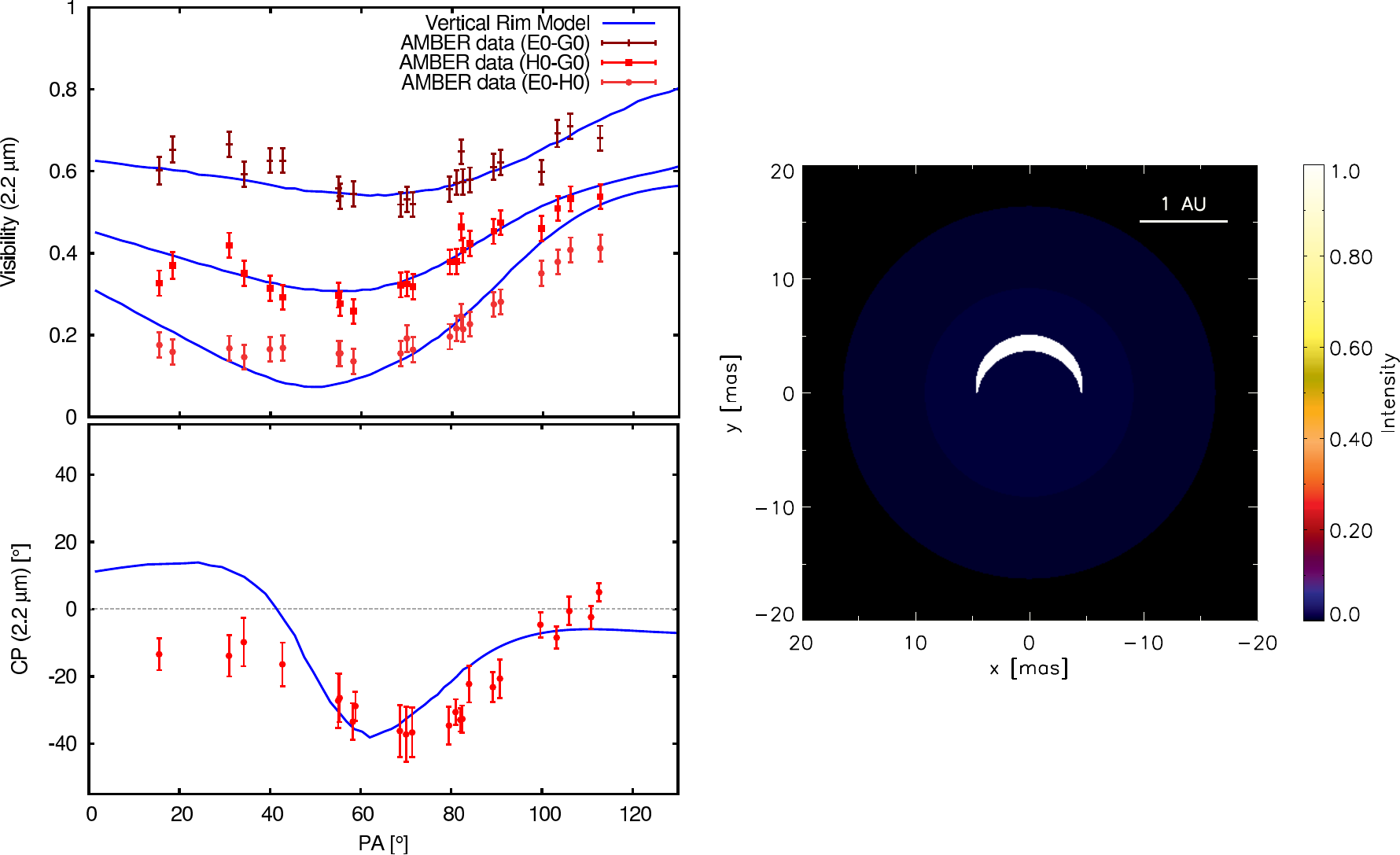}
  \caption{
    {\it Left:} Comparison between the AMBER observables and the model 
    visibilities ({\it top}) and closure phases ({\it bottom}) from
    our VERTICAL RIM model.
    {\it Right:} Image corresponding to our best-fit model.
  }
  \label{fig:modelVERTRIM}
\end{figure*}

Over the last decade, detailed physical models have been
developed to describe the 3-D structure of the inner dust rim
around Herbig~Ae stars.
\citet{dul01} provided a first mathematical description of the 
inner rim geometry and considered a perfectly vertical inner rim.
Since a vertical rim should result in very strong asymmetries,
it seems promising to test whether such a model can
explain the strong closure phases detected in our R\,CrA data.
To compute the brightness distribution, we assume that all
near-infrared flux emerges from the hot, illuminated rim surface.
In this model, the free parameters are the rim radius $R$,
the rim scale height $H/R$, the position angle $\phi$, 
the system inclination $i$, and the envelope parameters 
$\theta_{2}$ and $I_{2}/I_{1}$.
Our resulting best fit is shown in Fig.~\ref{fig:modelVERTRIM},
but provides no satisfactory representation of the data 
($\chi_r^2=3.25$).

\subsection{CURVED RIM model}
\label{sec:modelIN05}

\begin{figure*}[p]
  \centering
  \includegraphics[width=18cm]{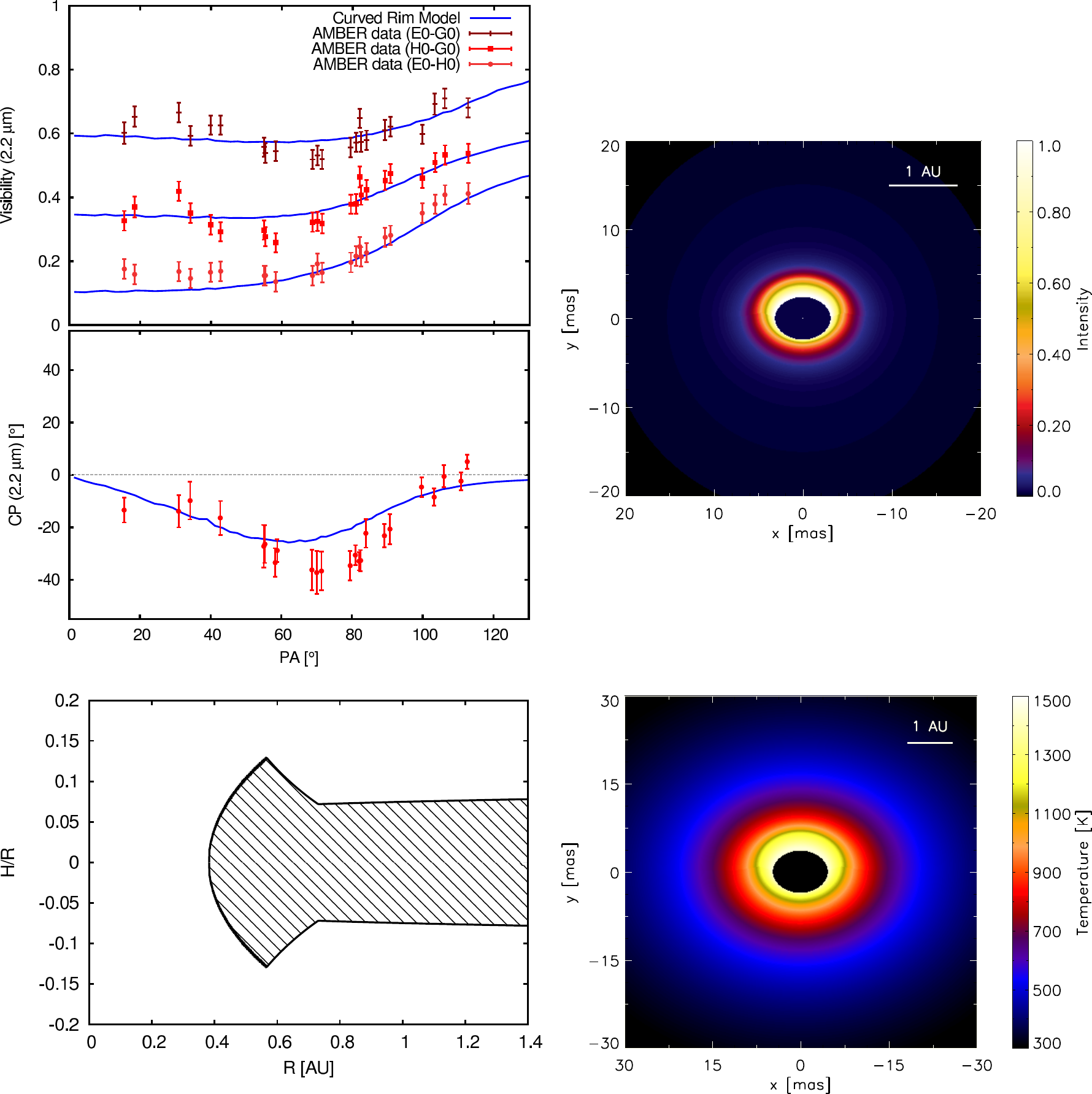}
  \begin{tabular}{c}
    \begin{minipage}[l]{9cm}
      \includegraphics[width=8.8cm,angle=0]{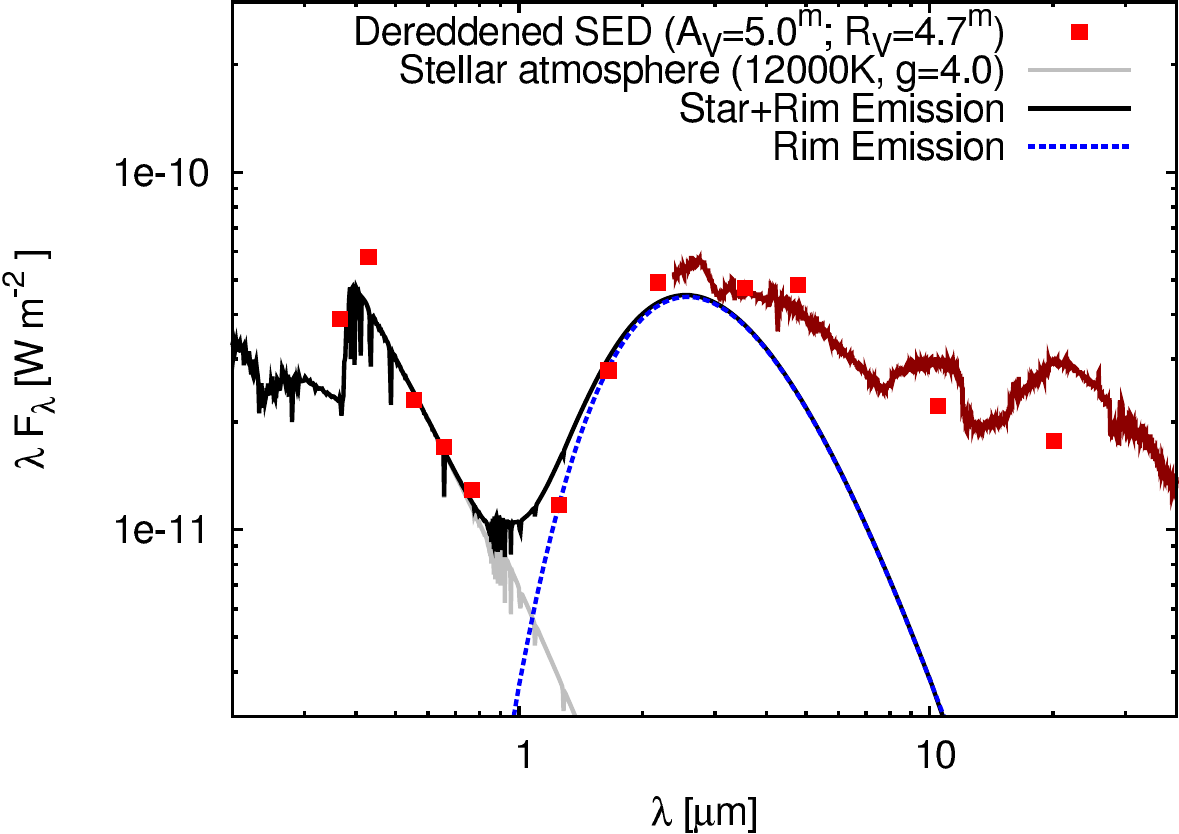} \\
    \end{minipage}
    \begin{minipage}[l]{9cm}
      \caption{
        {\it Top, left:} Comparison between the AMBER observables and the model 
        visibilities and closure phases from our CURVED RIM model.
        {\it Top, right:} Image corresponding to our best-fit model for the wavelength 2.2~$\mu$m.
        {\it Middle:} Photospheric height ({\it left}) and temperature distribution ({\it right}) 
        of the inner rim for our best-fit model.
        {\it Bottom, left:} SED of R\,CrA including photometric data from \citet[][ data points]{hil92} and
        an ISO spectrum (Frieswijk et al.\ 2007, SWS AOT-1 HPDP, dark-red curve).
        The grey curve shows the photospheric emission corresponding to a B8 model 
        atmosphere \citep{kur70} with luminosity 29~L$_{\sun}$, while the dashed blue curve
        shows the reprocessed light from the rim.
        The black curve represents the sum of both components.
      }
      \label{fig:modelIN05}
    \end{minipage}
  \end{tabular}
\end{figure*}

Following the pioneering work by \citet{nat01} and \citet{dul01}, several
studies aimed to refine the theoretical description of the rim geometry.
For instance, IN05 pointed out that the dust sublimation temperature 
should depend on the local gas density.
Accordingly, the dust sublimation temperature should be highest in the disk midplane 
and decrease with scale height $H$, resulting in a curved shape
of the inner rim.
In order to further increase the physical accuracy,
\citet{tan07} included the effect of dust grain sedimentation, showing that the
presence of a population of small and large dust grains results in an even stronger
rim curvature than predicted by the IN05 model.

To compute the radial dependence of the scale height of the disk
photospheric layer $H/R$, we follow the analytical approach by IN05.
The dependence of the dust sublimation temperature on the
local gas density $\rho_{g}$ (in g/cm$^{3}$) is described by the relation from
\citet{pol94}, namely $T_{\rm subl} = 2000~\mathrm{K} \cdot \rho_{g}^{0.0195}$.
In order to avoid unrealistic cutoffs at the outer edge of the puffed-up inner rim,
we include the more extended disk regions in our model as well and
construct our disk with the following components:
\begin{itemize}
\item[{\it (a)}] {\it Puffed-up inner rim:} This region extends from $R_{\rm subl}$
  to the point where the rim surface layer becomes optically thin to 
  the stellar radiation.
\item[{\it (b)}] {\it Shadowed region:} Entering the optically thin regime, the 
  rim can no longer maintain its puffed-up shape and the scale height 
  decreases as ${\rm d}(H/R)/{\rm d}R = -1/(8R)$, as derived by \citet{dul01}.
\item[{\it (c)}] {\it Flared outer disk:} At larger radii, the disk might enter
  a flared shape, which we parameterize with $H/R = H_{0} \left( R/R_{0} \right)^{9/8}$
  and $T(R) \propto R^{-3/4}$ \citep{ken87}.
\end{itemize}
In addition, the model includes the photospheric emission, 
which we estimate from the SED by comparing the stellar flux with the measured
near-infrared flux.
Please note that our near-infrared observations are mainly sensitive to regions 
{\it (a)} and {\it (b)}, while the parameters for the outer flared disk region are 
not constrained by our model fits.
Therefore, we fixed the parameters $H_{0}$ and $R_{0}$ to the typical values 
of $H_{0}=0.1$~AU and $R_{0}=10$~AU.
As total disk mass, we assume $M_{\rm disk}=0.012~M_{\sun}$ \citep{gro07} 
with a standard gas-to-dust ratio of 100:1.
Besides the luminosity $L$ which is incident
on the disk rim, the most sensitive parameter in the 
IN05 model is $\epsilon$.
This variable is defined as the ratio of the Planck mean opacity at the dust 
evaporation temperature $T_{\rm subl}$ to the Planck mean opacity at the stellar effective 
temperature $T_{\star}$ and defines the cooling efficiency
of the dust grains.  It therefore depends on the precise dust chemistry and
grain size distribution.
For a given grain population, $\epsilon$ typically increases for larger 
grain sizes, moving the rim location closer to the star (see IN05).
Above a certain threshold ($\epsilon \gtrsim \epsilon_{\rm cr}=1/\sqrt{3}$), 
the location of the optically thick dust rim stays nearly constant, 
and the largest grains can only survive closer to the star in a 
optically thin regime.  Therefore, we imposed for our fit that
$\epsilon$ cannot exceed $\epsilon_{\rm cr}$. In addition, 
the model assumes that absorption and continuum emission 
by gas inside the dust sublimation radius is negligible.
This assumption is supported by the theoretical work of \citet{muz04},
who found that for the accretion rate of R\,CrA 
\citep[$\dot{M}_{\rm acc}=10^{-7.12}~M_{\sun}/{\rm yr}$,][]{gar06}, the gas 
located inside the dust sublimation radius should be optically thin.

After computing the disk scale height $H(r)$ and blackbody temperature $T(r)$
for each disk annulus, we construct synthetic images of the 
disk brightness distribution.
By computing these images for each spectral channel separately, 
we can simulate the wavelength-differential changes resulting
from the disk temperature distribution.

The free disk model fitting parameters are the position angle $\phi$, the system inclination $i$, 
the grain cooling efficiency $\epsilon$, and the envelope parameters $\theta_{2}$ and $I_{2}/I_{1}$.  
Of course, another crucial parameter is the luminosity $L$, 
which is heating the inner dust rim, and which might contain stellar light contributions ($L_{\star}$) 
as well as contributions due to active accretion ($L_{\rm acc}$).
Although the precise amount of either of these contributions is still not well known, 
it is likely that the total incident luminosity $L := L_{\star} + L_{\rm acc}$
is neither as low as $\sim 2~$L$_{\sun}$ (corresponding to the stellar luminosity 
of a F5 star), nor that the full bolometric luminosity of 
$\sim 99~$L$_{\sun}$ \citep{bib92} derived from KAO photometry
can be solely attributed to this single source.
Therefore, we treat $L$ as a free parameter and keep the other
stellar parameters, which have a less significant influence on our modeling results,
fixed to the values corresponding to a B8-type star
\citep[$T_{\rm eff}=12,000$~K, $M_{\star}=3.0~M_{\sun}$,][]{bib92}.
Then, we adjust the absorption coefficient $A_V$ to roughly reproduce 
the SED of R\,CrA at UV/visual wavelengths.
Following \citet{bib92}, the total-to-selective extinction ratio $R_V$ was fixed to 4.7.
In order to include the emission of the puffed-up inner rim in our model SED,
we follow the approach outlined by IN05, assuming that the rim surface layer reprocesses the incident 
light and re-emits it as a blackbody of temperature $\sim 1400$~K.
When modeling the SED of R\,CrA, one faces the
general problem that the visual/near-infrared emission also contains 
major scattered flux contributions from the circumstellar envelope, 
as described in Sects.~\ref{sec:model2GAUSS} and \ref{sec:discussionIntProfile}.
Due to this strong "contamination", we decided not to include the
SED as an additional modeling constraint, but to treat the fraction of
stellar luminosity which is reprocessed at near-infrared wavelengths
$\beta := L_{\rm NIR}/L$ as a free model parameter, which is adjusted 
to match the measured SED of R\,CrA.
In Sect.~\ref{sec:discussionRimGeometry}, we will compare the derived 
$\beta$-value with the predictions from the IN05 model 
and discuss the consistency between the measured SED 
and the puffed-up inner rim scenario.

Using our modeling approach, we find that the parameter combination
$L=29$~L$_{\sun}$ and $\epsilon=\epsilon_{\rm cr}$
is able to simultaneously reproduce the measured visibilities and closure phases 
($\chi_r^2=2.14$, Fig.~\ref{fig:modelIN05}, {\it top, left}) as well as the
visual- to near-infrared SED (with $A_V=5.0$ and $\beta=0.35$; see Fig.~\ref{fig:modelIN05}, {\it bottom, left}).
For the following reasons, both parameters appear to be relatively well constrained:
\begin{itemize}
\item Assuming a higher value for $L$ would push the dust sublimation radius
  outwards, resulting in too low visibilities 
  (e.g., for $L=50$~L$_{\sun}$, the best-fit $\chi^2_r$-value increases already to 2.85).
\item A lower luminosity would require to increase $\beta$ above physically
  reasonable values (e.g., for $L=20$~L$_{\sun}$, already more than 60\% of the stellar light 
  would have to be reprocessed by the inner rim).
\item $\epsilon$-values significantly lower than $\epsilon_{\rm cr}$ seem to be excluded, 
  since this would again result either in too low visibilities,
  or require to increase $\beta$ substantially.
\end{itemize}
We conclude, that already with a relatively small number of free parameters
($L$, $\epsilon$, $\phi$, $i$, $\theta_{2}$, $I_{2}/I_{1}$), it is possible
to simultaneously reproduce the interferometric observables ($\chi_r^2=2.14$) and 
the relevant parts of the SED.  At wavelengths $\gtrsim 3~\mu$m, our model is not able to 
reproduce the SED, since we do not include the flux contributions from the 
more extended disk regions nor from the circumstellar envelope.
In Fig.~\ref{fig:modelIN05} we show the model visibilities and closure phases, the SED, 
as well as the computed photospheric scale height $H/R$ and the temperature distribution
for our best-fit model.  In this model, the inner rim radius is located at
0.38~AU.

\begin{figure*}[tbp]
  \centering
  \includegraphics[width=18cm,angle=0]{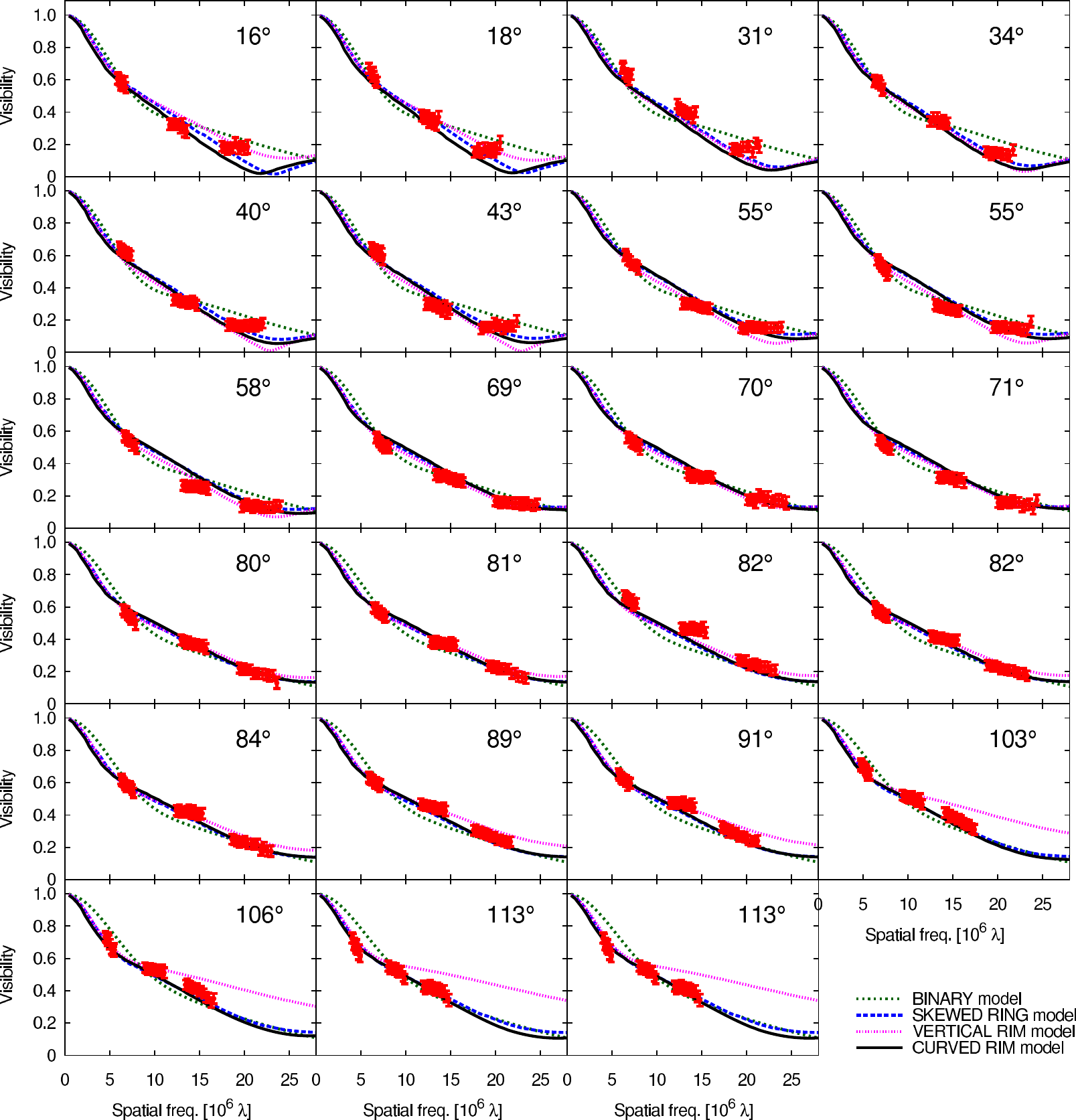} \\
  \caption{Comparison of the measured wavelength-dependent
    AMBER visibilities and the model visibilities corresponding to the 
    best-fit model assuming our
    BINARY model (Sect.~\ref{sec:modelbinary}), 
    SKEWED RING model (Sect.~\ref{sec:modelskewedring}), 
    VERTICAL RIM model (Sect.~\ref{sec:modelVERTRIM}), and 
    CURVED RIM model (Sect.~\ref{sec:modelIN05}).
  }
  \label{fig:overviewVIS}
\end{figure*}

\begin{figure*}[tbp]
  \centering
  \includegraphics[width=18cm,angle=0]{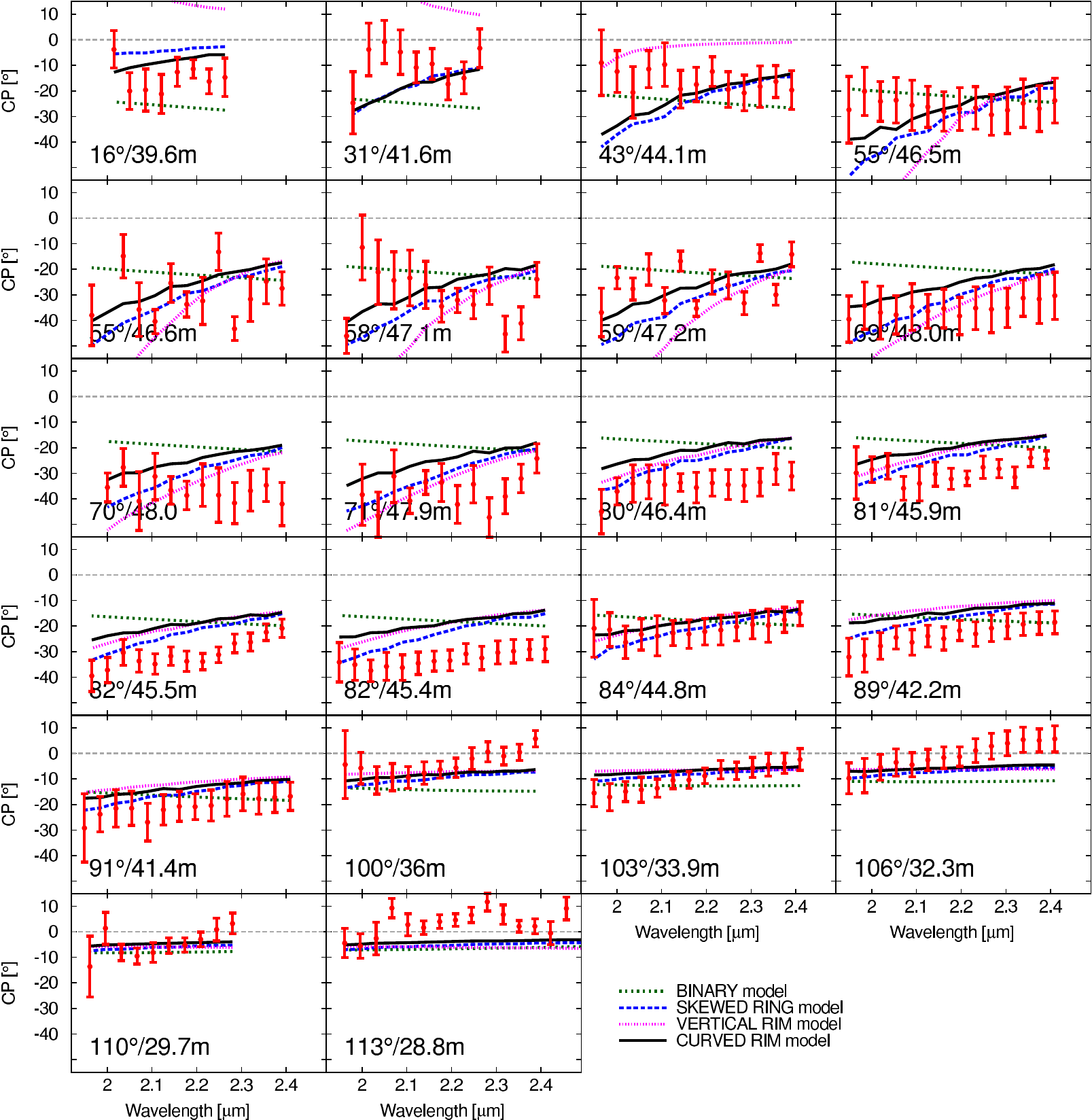} \\
  \caption{Comparison of the measured wavelength-dependent closure phases
    with the predictions from our best-fit
    BINARY model (Sect.~\ref{sec:modelbinary}), 
    SKEWED RING model (Sect.~\ref{sec:modelskewedring}), 
    VERTICAL RIM model (Sect.~\ref{sec:modelVERTRIM}), and 
    CURVED RIM model (Sect.~\ref{sec:modelIN05}).
    In each panel, the position angle of the three baselines and 
    the projected length of the longest baseline $B_{\rm E0-H0}$ 
    is given.  Given the equal spacing of the employed linear telescope
    array, the other two baselines have projected length of 
    $B_{\rm E0-G0}=B_{\rm E0-H0}/3$ and $B_{\rm G0-H0}=2 B_{\rm E0-H0}/3$.
  }
  \label{fig:overviewCP}
\end{figure*}

\section{Discussion}
\label{sec:discussion}

\subsection{Radial intensity profile}
\label{sec:discussionIntProfile}

As shown in Sect.~\ref{sec:modelspherical}, it is evident that 
the radial intensity profile measured toward R\,CrA cannot be
well represented with simple one-component geometries, 
such as rings.
Instead, the visibility function has an approximately 
linear slope (Fig.~\ref{fig:visHKvisPA}, {\it left}), showing 
remarkable similarities with the visibility profile measured 
by \citet{ack08} on the much more massive 
YSO MWC\,297.  In both cases, the visibilities could be 
approximated reasonably well with two Gaussian components.
For the case of R\,CrA, we interpret this finding with
the presence of an extended optically thin envelope, which contributes 
about one-third of the total $K$-band flux, likely through scattered light.
This extended emission might represent scattered light from 
an optically thin spherical halo
or from the walls of an associated outflow cavity.
The presence of an envelope around R\,CrA was already 
deduced earlier from SED fits \citep{nat93}
and from polarimetric observations \citep{cla00}
and seems in line with the early evolutionary stage of the object.

Independent of the precise physical interpretation of the 
measured visibility profile, our finding shows that
the commonly applied ring model fits often represent an
over-simplication of the complex environment around YSOs.
Since these ring fits are commonly applied to 
single-baseline, broad-band interferometric observations,
the uncertainties of the real underlying radial intensity
profile likely contribute significantly to the 
scatter which is observed in size-luminosity diagrams \citep{mon05}.
Therefore, very detailed interferometric studies on some
additional sources will be essential to identify
the real underlying source structure and
will have a direct impact on the proper interpretation of 
the data obtained in survey-type observations.

\subsection{Non-detection of the proposed binary companion}
\label{sec:discussionNoBinary}

Given the earlier speculations about the existence of a companion 
star for R\,CrA \citep[][see discussion in Sect.~\ref{sec:rcra}]{tak03,for06}, 
it is an important result that we did not find
indications for binarity of R\,CrA from our study.
As discussed in Sect.~\ref{sec:modelbinary}, it is possible
to rule out the presence of wide separation binaries
($\rho \gtrsim 40$~mas) with rather general arguments.
To probe for binaries with smaller separation, we constructed
geometric models, which also take into account that each
star might be associated with a circumstellar disk (represented by 
Gaussians in our model).
Since this model provided only a poor representation of our data, 
we consider such a close binary scenario rather unlikely.
Furthermore, the best-fit parameters of the binary model
do not correspond to a physical solution, since the extension of
the Gaussian components (1.9 and 0.8~AU) exceeds even their
projected separation (0.8~AU), 
resulting in an overlap of the two components.
Assuming that the physical separation is of similar order as the
projected separation, a hypothetical binary on such small spatial 
scales would not be stable and would quickly disrupt the 
circumstellar disks around both components.

Of course, with the existing data, it is difficult to rule out 
scenarios which might involve three or more spatial components,
such as, for instance, a binary system with circumstellar
and circumbinary disks.
Also, our observations do not rule out the existence of a
very deeply embedded companion, which might not contribute
significantly to the near-infrared emission.
However, this scenario is rather unlikely,  
since the SED of R\,CrA peaks in the $K$-band
(Fig.~\ref{fig:modelIN05}, {\it bottom, left}), which makes the existence of an 
additional strong, far-infrared-emitting component unlikely.

Our non-detection of a binary component around R\,CrA
suggests that the detected hard X-ray emission is not associated
with a hypothetical Class~I source \citep{for06}, but with 
a single Herbig~Ae star.
This finding on R\,CrA is in line with the study
by \citet{ham08}, who concluded from statistical arguments
that the X-ray emissions of Herbig~Ae/Be stars have an intrisic origin.
Since neither coronal nor shock-excited X-ray emission is
expected for A-type stars, more theoretical work is 
clearly required in order to identify the X-ray-emitting mechanism of 
single intermediate-mass pre-main-sequence stars such as R\,CrA.

\subsection{Constraints on the rim geometry}
\label{sec:discussionRimGeometry}

\begin{figure*}[tbp]
  \centering
  \includegraphics[width=18cm,angle=0]{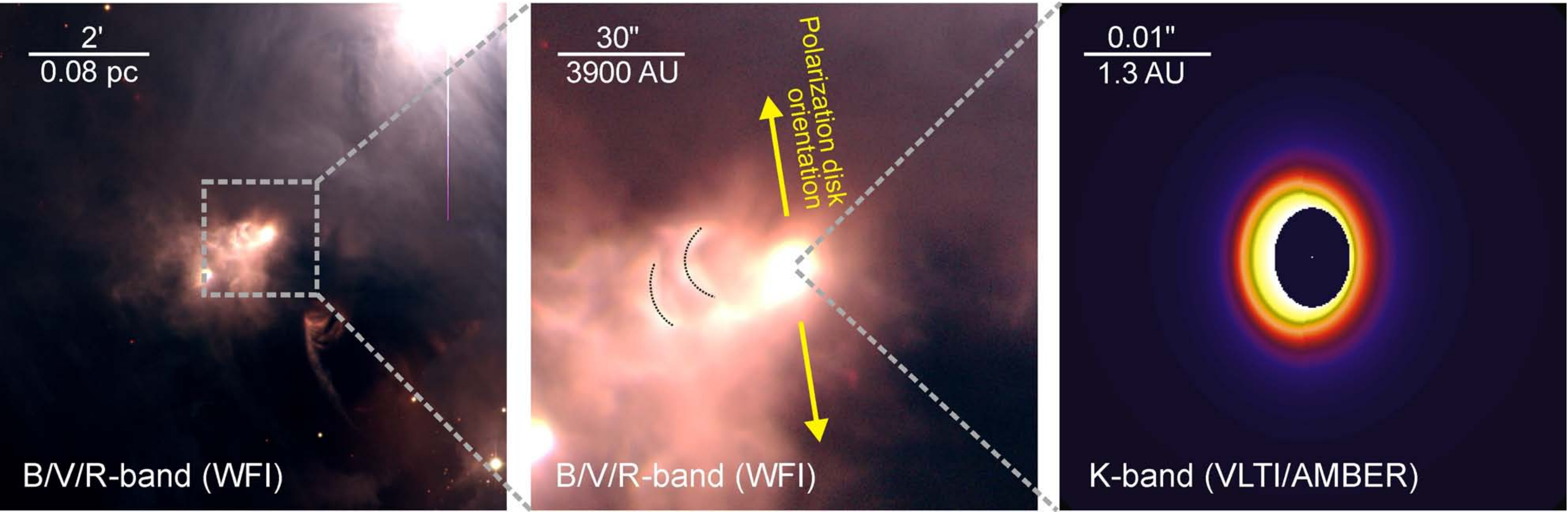} \\
  \caption{
    At visual wavelengths the field around 
    R\,CrA is dominated by the reflection nebula NGC\,6729 
    ({\it left} panel; color composite: blue: $B$-band, green: $V$-band, red: $R$-band;
    North is up and east is to the left).
    Within the reflection nebula, two bow shock-like structures appear
    ({\it middle} panel, dashed lines), suggesting an east-western outflow axis,
    which is roughly perpendicular to the polarization
    disk reported by \citet{war85}
    and the sub-AU disk resolved by our VLTI/AMBER observations 
    ({\it right} panel; CURVED RIM model image; 
    please note that there is still an 180{\deg}-ambiguity in the 
    derived orientation, as discussed in Tab.~\ref{tab:modelfitting}).
    The $B$/$V$/$R$-band images were taken with the Wide Field Imager (WFI)
    at the 2.2-m~MPG/ESO telescope on La~Silla 
    (ESO Press Photo 25a-b/00; image courtesy: 
    European Southern Observatory and F.~Comeron).
  }
  \label{fig:overview}
\end{figure*}

Given that our best-fitting results were obtained with the 
skewed ring model (Sect.~\ref{sec:modelskewedring}) 
and the curved rim model (Sect.~\ref{sec:modelIN05}),
we consider it most likely that our observations are 
directly tracing the asymmetries introduced by material at the
inner truncation radius of the dust disk around R\,CrA.
This interpretation is also supported by the 
fact that the $H$- and $K$-band visibilities follow 
within the error bars the same visibility profile 
(Sect.~\ref{sec:observations} and Fig.~\ref{fig:visHKvisPA}, {\it left}),
providing important information about the temperature distribution
of the near-infrared emitting disk material.
Earlier spectro-interferometric observations
on two other Herbig~Ae/Be stars \citep{kra08a,ise08} 
measured in the $H$-band a significantly higher visibility 
than in the $K$-band, which was interpreted as the presence 
of an optically thick, hot gaseous disk component 
located inside of the dust sublimation radius.
For R\,CrA, we do not measure this visibility increase towards
shorter wavelengths, indicating that the disk material in the
probed temperature range ($\gtrsim 1000$~K) 
is located at similar distances from the star, 
likely in a narrow region around the dust sublimation radius.

Concerning the detected asymmetries, it is interesting to compare
our results with earlier studies on 
HD\,45677 \citep[$\Phi \lesssim 27${\deg},][]{mon06} and
\object{AB\,Aur} \citep[$\Phi \lesssim 4${\deg},][]{mil06b}, which also
detected asymmetries in the inner region around these Ae- and Be-type stars.
Comparing HD\,45677 and R\,CrA, it is remarkable that the closure phases
from the two objects could be represented well with a skewed ring model.
However, a major difference between both objects concerns the
evolutionary stage: HD\,45677 is likely an evolved B[e] star, while
R\,CrA is a young, actively accreting Herbig~Ae star.
Compared to the Herbig~Ae star AB\,Aur, a major difference 
concerns the physical nature of the asymmetry.
For AB\,Aur, the modeling of \citeauthor{mil06b} showed that
the asymmetries around this object originate from a localized
region within the disk (maybe a hot accretion spot), while
our data seems more consistent with a diffuse structure
around the dust sublimation radius.
Another Herbig~Ae/Be star where strong asymmetries were detected
is LkH$\alpha$101. 
Near-infrared aperture-masking observations by \citet{tut01} 
revealed a skewed circular structure at a distance of $\sim 3.4$~AU 
around this rather massive ($M_{\star} \approx 5-10~M_{\sun}$) 
and luminous ($L_{\star}\approx 480~L_{\sun}$) Herbig~Be star.
Remarkably, the structure in their observations shows some 
similarities with our best-fit skewed ring model, which might
suggest that, contrary to earlier suggestions \citep{vin07},
Herbig~Ae and Herbig~Be have a common rim structure.

To determine the precise rim shape, we have fitted three different
models, namely the skewed ring model (Sect.~\ref{sec:modelskewedring}), 
the vertical rim model (Sect.~\ref{sec:modelVERTRIM}), and 
the curved rim model (Sect.~\ref{sec:modelIN05}).
Formally, the best agreement was found with the geometric
skewed ring model ($\chi_r^2=1.96$), although
it should be noted that the skewed ring model
is a purely geometric model and has the 
largest number of adjustable free parameters
(5~disk~\&~2~envelope parameters).
On the other hand, the curved rim model ($\chi_r^2=2.14$)
has less free parameters 
(3~disk, 2~envelope, and 1~stellar parameter) 
and is based on detailed theories about disk structure and dust properties,
allowing a direct physical interpretation of the
fitted parameters.
In this context, the dust cooling efficiency parameter $\epsilon$ 
is of particular interest, since this parameter is directly
related to the dust properties in the inner disk regions.
For instance, \citet{ise06} used the curved rim model to
derive the maximum dust grain size for a sample
of five Herbig~Ae/Be stars, finding in all cases relatively 
large dust grains ($\gtrsim 0.2~\mu$m).
Assuming the same Silicate dust chemistry, we derive for R\,CrA 
dust grain sizes larger than $1.2~\mu$m (corresponding to $\epsilon \gtrsim \epsilon_{\rm cr}$),
indicating that the dust in the inner disk regions was already significantly processed by grain growth.
In order to reproduce the interferometric data and the SED simultaneously,
we require that the incident luminosity, which is heating the inner dust rim,
is around $29~$L$_{\sun}$ and that $35$\% of the stellar light 
is reprocessed at near-infrared wavelengths.
This fraction is somewhat
higher than the theoretical value of $\sim 20$\% predicted by the IN05 model (see Fig.~4 in IN05). 
However, this might reflect the fact that a significant fraction of the
near-infrared emission is likely not reprocessed light from the inner rim, 
but scattered light contributions from a circumstellar envelope 
(Sects.~\ref{sec:model2GAUSS} and \ref{sec:discussionIntProfile}).
Although it is currently unfortunately not possible to properly decompose the 
disk and envelope SED,
it is likely that the $\beta$-value has to be corrected by a similar factor as the
determined envelope/disk $K$-band flux ratio of 1/3 (Sect.~\ref{sec:model2GAUSS}),
which would result in a good agreement with the expected theoretical value.
Alternative explanations for the derived high value of $\beta$ would include
an underestimation of the gas disk mass ($M_{\rm disk}=0.012~M_{\sun}$) and the 
resulting underestimation of the Silicate sublimation temperature
or the presence of highly refractory dust species which sublimate
at higher temperature than Silicate grains.

For the derived luminosity of $\sim 29~$L$_{\sun}$, we consider that the
dominant contribution is photospheric emission, while the contributions from
active accretion are likely less significant, as indicated by the observed
low Br$\gamma$-line luminosity \citep{gar06}.
Accordingly, only about one-third of the total bolometric luminosity
\citep[$\sim 99~$L$_{\sun}$,][]{bib92} can be attributed to R\,CrA, while 
the remaining fraction is likely due to contamination 
from other, more deeply embedded members of the Coronet cluster 
in the large ($\sim$45\arcsec) field-of-view of the KAO satellite.

For the disk inclination angle, we consider the value determined with
the curved rim model ($i=35${\deg}) our most reliable estimate
\footnote{In particular, this intermediate inclination angle seems more reliable 
than the very low inclination of $14${\deg} determined with the 
skewed ring model, reflecting the fact that in the skewed ring model
the dominant free parameter is the skew parameter $s$, while 
the inclination $i$ has only a minor effect on the model appearance, 
and therefore, this parameter is only poorly constrained.}
and note that the value is also in good agremeent
with the inclination angle of $\sim 40${\deg}, which was
derived by \citet{cla00} for the reflection nebula NGC\,6729 
using polarimetric observations.
NGC\,6729 also shows some remarkable fine-structure,
  in particular two bow shock-like features \citep[labeled E and F in][]{cla00},
  which appear east of R\,CrA (Fig.~\ref{fig:overview}, {\it left} and {\it middle}).
  As already suggested by \citeauthor{cla00}, these structures
  might be created by vigorous, periodic outflow activity from R\,CrA.
  The sub-AU scale disk structure resolved by our AMBER observations
  is oriented approximately perpendicular to these bow shocks 
  ($\theta=180,190${\deg} for the CURVED RIM model and the SKEWED RING model, respectively),
  suggesting that R\,CrA is the driving engine which has
  created these bow shocks (Fig.~\ref{fig:overview}, {\it right}).
  The derived disk orientation is also consistent with the orientation 
  of the polarization disk reported by 
  \citet[][ $\theta=189 \pm 5${\deg}; see Fig.~\ref{fig:overview}, {\it middle}]{war85}.
  The bright feature extending from R\,CrA to the south-east \citep[labeled C in][]{cla00} was
  identified by \citeauthor{cla00} as scattered light from the walls of a parabolic outflow cavity.

When judging the quality of the obtained fit, it is 
interesting to note that the IN05 fitting results improved slightly
by taking the expected disk temperature gradient effects into account
($\chi_r^2=2.14$). 
For comparison, when we simulate a monochromatic brightness
distribution ($\lambda=2.2~\mu$m), we yield $\chi_r^2=2.38$.
Nevertheless, with a $\chi_r^2$-value of 2.14, 
the quality of the best fit IN05 model is still relatively poor, 
indicating that the rim geometry might still not be adequately 
represented by the employed rim model.
Therefore, it will be important to investigate in future studies
how the models can be modified to yield a better 
representation of the measured visibilities and closure phases.
For instance, including dust sedimentation and grain growth 
effects should result in an even more curved rim shape and
a broader near-infrared emitting rim region than predicted 
by the current model \citep{tan07}. 
Based on the good fitting results obtained with a 
skewed ring model with a large fractional ring width of $0.8$
(Sect.~\ref{sec:modelskewedring}), we expect 
that such a modification might improve the visibility/phase fit.
Possibly, an improvement could also be obtained by including the 
optically thin dust condensation zone which is expected to extend
inwards of the dust rim, yielding a more diffuse
rim structure \citep{kam09}.
Other very promising modifications concern the presence of 
multiple grain species, in particular highly refractory metal oxides such 
as iron or corundum, and a consistent treatment of backwarming effects.
\citeauthor{kam09} have shown that these effects can move the
dust sublimation radius nearly by a factor of~2 closer to the star.
With our fitting procedure, this change would result in a higher
derived stellar luminosity 
(i.e.\ more consistent with the measured bolometric luminosity) 
and a lower fraction of reprocessed light
(i.e.\ closer to the theoretical expectation).
Thus, we expect that these model refinements aim in the
right direction in order to further improve the physical consistency
of the puffed-up inner rim scenario for R\,CrA,
although a detailed data modeling study will be required
in order to test whether the model refinements
also affect the rim morphology, as required to yield
a better $\chi^2$-fit.
Finally, one should take into account that our observations
probe relatively small physical scales in a highly dynamical 
and complex environment.
Therefore, it cannot be excluded that our interferometric observations
might also be affected by local inhomogenities or 
astrophysical processes, which are not yet completely included
in the employed disk rim models, such as active accretion or 
the influence of outflow launching on the disk structure.

\section{Conclusions}
\label{sec:conclusions}

We summarize our findings as follows:
\begin{itemize}
\item Using 24 VLTI long-baseline interferometric measurements, 
  we studied the near-infrared emission from the Herbig~Ae star R\,CrA 
  and could clearly resolve the inner circumstellar 
  environment in the $H$- and $K$-band. 
  Besides this spectral coverage, our observations also provide 
  a good sampling of the visibility function ($0.8 \lesssim V \lesssim 0.1$) 
  towards a wide range of position angles (PA=16-113{\deg}).
\item Even when considering only a single position angle, 
  the measured visibility profile cannot be represented with the commonly
  applied one-component models, but indicates a more complex object geometry,
  likely including a disk and an envelope component.
\item The measured closure phase signals (up $\Phi \sim 40${\deg}) clearly
  indicate a strongly asymmetric brightness distribution.
\item We find that the detected asymmetries are likely not related to the presence
  of a companion star, but directly trace the vertical structure of the 
  disk around the dust sublimation radius ($R_{\rm subl} \sim 0.4$~AU).
  When seen under intermediate inclination,
  the increased disk scale height causes obscuration and shadowing effects,
  which result in the observed asymmetries.
  To constrain the precise rim geometry, we tested three
  geometric and physical models, including the generic skewed ring model
  and rim models with a vertical and a curved rim shape. 
  Clearly, models with a smooth brightness distribution 
  (i.e.\ the skewed ring model and the curved rim model) 
  provide a much better
  representation of our data than the sharp rim structure predicted by
  the vertical rim model.
\item Confronting our data with the detailed mathematical description 
  of the rim structure presented by IN05, we find that we can 
  reasonably well reproduce the measured SED, as well as the 
  visibilities and closure phases
  with a disk inclination angle of 35{\deg} and an incident
  luminosity of 29~L$_{\sun}$.
  Given the absence of indications for strong active accretion, we consider
  that this luminosity is mainly of photospheric origin.
  For the dust grain size, we find that the presence of relatively large 
  Silicate dust grains ($\gtrsim 1.2~\mu$m) is required to obtain agreement
  between the model and our data.  
  The derived disk position angle of $\sim 180-190${\deg} agrees well with the 
  orientation of the polarization disk ($\theta=189 \pm 5${\deg}).
  Perpendicular to the disk axis, two bow shocks
  appear in the associated reflection nebula NGC\,6729,
  suggesting that the detected disk is driving an outflow, 
  which has shaped the bow shock-like structures.
\end{itemize}

Presenting one of the most comprehensive high-angular resolution studies 
on the inner structure of YSO disks, our findings provide strong evidence
for the existence of a curved, puffed-up inner dust rim in Herbig~Ae stars.
However, additional work is needed to further constrain the detailed
rim geometry and to distinguish the rim morphology from
local brightness inhomogenities, which might, for instance, be caused
by hot accretion spots, spiral disk density patterns, 
clumpiness, or other transient phenomena.
Given the complexity one likely faces on the 
probed sub-AU scales, this task will urgently require the model-independent 
aperture synthesis imaging capabilities which are just becoming available
for various infrared interferometric facilities.

\begin{acknowledgements}
We thank the ESO Paranal team for their efforts and 
excellent support during our visitor mode observations and 
L.~Testi for constructive comments, which helped to improve this paper.
\end{acknowledgements}

\bibliographystyle{aa}
\bibliography{12990}

\end{document}